\pdfoutput=1
\documentclass[twocolumn,aps,prb,showkeys,superscriptaddress,letterpaper,floatfix]{revtex4-2}
\usepackage{amsmath}
\usepackage{amssymb}
\usepackage{graphicx}
\graphicspath{{plots/}}
\usepackage[dvipsnames]{xcolor}
\usepackage{bm}
\usepackage{subcaption}
\usepackage{ragged2e}
\usepackage{caption}
\makeatletter
\def\@makesimplecaption#1#2{%
  \setbox\@tempboxa\hbox{\small #1. #2}%
  \ifdim\wd\@tempboxa>\hsize
    {\small #1. #2\par}
  \else
    \hbox to\hsize{\hfil\box\@tempboxa\hfil}
  \fi
}

\long\def\@makecaption#1#2{%
  \par\vskip\abovecaptionskip
  \begingroup
    \small
    \justifying\noindent 
    \@makesimplecaption{#1}{#2}%
  \endgroup
  \par\vskip\belowcaptionskip
}
\makeatother
\usepackage{chemfig}

\usepackage{tikz}
\usepackage{pgfplots}
\usepgfplotslibrary{groupplots}
\pgfplotsset{compat=newest}

\usepackage{comment}
\renewcommand{\mathbf}{\bm}

\usepackage[unicode=true,
 bookmarks=true,bookmarksnumbered=false,bookmarksopen=false,
 breaklinks=false,pdfborder={0 0 0},pdfborderstyle={},backref=false,colorlinks=true,]
 {hyperref}
\hypersetup{linkcolor=blue,citecolor=blue,urlcolor=blue}

\DeclareMathOperator{\dn}{dn}
\DeclareMathOperator{\sn}{sn}
\DeclareMathOperator{\am}{am}
\DeclareMathOperator{\sech}{sech}

\DeclareMathOperator{\sign}{sgn}
\DeclareMathOperator{\hc}{h.c.}

\allowdisplaybreaks[1]

\usepackage{xcolor} 
\definecolor{C0}{RGB}{88,80,141}
\definecolor{C1}{RGB}{188,80,144}
\definecolor{C2}{RGB}{255,99,97}
\definecolor{C3}{RGB}{255,166,0}

\begin{document}

\title{Electrostatically induced topological phase transitions in polyacetylene molecules}

\author{Tomás Suleiman}

\affiliation{Instituto de Física La Plata - CONICET, Diag 113 y 64 (1900) La Plata, Argentina}

\author{Aníbal Iucci}

\affiliation{Instituto de Física La Plata - CONICET, Diag 113 y 64 (1900) La Plata, Argentina}
\affiliation{Departamento de Física, Universidad Nacional de La Plata, cc 67, 1900 La Plata, Argentina.}

\author{Alejandro M. Lobos}

\affiliation{Facultad de Ciencias Exactas y Naturales, Universidad Nacional de Cuyo and CONICET, 5500 Mendoza, Argentina}
\affiliation{Instituto Interdisciplinario de Ciencias Básicas (CONICET-UNCuyo)}

\begin{abstract}
We study the electronic properties of a linear trans-polyacetylene (tPA) molecule capacitively coupled to an external gate voltage $V_g$ of width $d$. We describe this system using the Takayama-Lin-Liu-Maki model, a continuum version of the Su-Schrieffer-Heeger model, and analyze it within the Abelian bosonization formalism, which allows us to treat both electronic and lattice degrees of freedom and to incorporate the effects of repulsive Coulomb interactions among electrons. The global ground state describing simultaneously the electronic charge-density field as well as the lattice dimerization field of a tPA molecule is shown to consist of multikink solutions of a modified sine-Gordon equation for the charge-density field, which is controlled by $V_g$, the width $d$, and the Luttinger parameter $K$ encoding the strength of electron-electron interactions. We show that these solutions belong to distinct topological sectors labeled by an integer invariant $q$ that simultaneously quantifies both the bound charge and the number of domain walls in the dimerization pattern induced at the gated region. Increasing $V_g$ drives a sequence of topological phase transitions characterized by abrupt changes in  $q$. We further examine the effect of repulsive Coulomb interactions on the resulting topological phase diagram, and, finally, we discuss the relevance of our findings for potential nanoelectronic devices based on gated tPA molecules. 
\end{abstract}

\maketitle

\section{Introduction}

Trans-polyacetylene (tPA), a linear chain of carbon atoms with alternating long and short bonds, was the first organic polymer found to exhibit conductive properties \cite{Heeger1988a, Barford13_Electronic_and_Optical_Properties_of_Conjugated_Polymers, Farchioni_Grosso_Organic_electronic_metals_book}. Its Peierls-dimerized ground-state structure hosts topological solitons  bound to lattice deformations called domain walls (DWs). These excitations were predicted theoretically in the late 1970s in the framework of the celebrated Su-Schrieffer-Heeger (SSH) model \cite{Su79_Solitons_in_polyacetylene, Su80_Soliton_excitations_in_PA,Heeger1988a} and were experimentally confirmed  via optical spectroscopy and magnetic resonance techniques \cite{photoinduced, photoexcitations, goldberg1979electron, weinberger1980electron}. From a field-theoretical perspective, solitons in tPA are a condensed-matter realization of the celebrated Jackiw-Rebbi soliton excitation with fractional charge $e/2$ \cite{Jackiw76_Jackiw_Rebbi_soliton}.

Despite the fact that solitons in tPA have been very well studied, recent progress in nanofabrication techniques has generated renewed interest in these systems. In particular, on-surface synthesis of molecular tPA chains on metallic Cu(110) \cite{Wang19_Solitons_in_individual_PA_molecules} and other related $\pi$-conjugated polymers on Au(111) \cite{cirera2020tailoring} surfaces, have enabled to address individual molecules via atomically-resolved scanning tunneling microscopy (STM) \cite{Wang19_Solitons_in_individual_PA_molecules}. Based on these experimental advances, recent works have put forward new strategies to harness the topological nature of the elementary excitations of tPA [and other related one-dimensional (1D) compounds] for their use in novel electronic nanodevices \cite{HernangomezPerez20_Solitonics_with_PA, park2022creation, Arancibia25_Towards_electrical_DW_control}. In particular,  the possibility to induce controllable DWs in a single tPA molecule via capacitively-coupled gate voltages was studied in Ref. \cite{Arancibia25_Towards_electrical_DW_control}. 
Interestingly, that work showed that by locally breaking the particle-hole symmetry at the position of the gate, the number of the DWs in the molecule and the charge induced at the gate could be externally modified by the applied gate voltage $V_g$. At certain critical values $V^{(q)}_g$ (with $q>1$), the device is predicted to undergo a topological transition between ground states with $q-1$ and $q$ DWs at which the lattice configuration abruptly changes, and  the excess charge bound at the region of the gate concomittantly jumps from $-(q-1)e$ to $-qe$ \cite{Arancibia25_Towards_electrical_DW_control}. At either side of $V^{(q)}_g$, the induced charge at the gate remains robustly quantized, a fact that could be relevant in novel designs of nanoelectronic devices based on topology. 

From a different perspective, the interplay of  topology and strong correlations is a theoretically challenging issue which has attracted a lot of interest in condensed-matter physics in the last decades. The emergence of topological phases driven by strong interactions remains one of the most intensively explored areas of research.  However, in contrast with topological band theory \cite{Kitaev_TI_classification, Ryu10_Topological_classification}, which allows a symmetry-based classification of different topological classes of noninteracting gapped fermionic systems, at present there is a lack of consensus regarding a unified theoretical picture for strongly interacting fermions, and different frameworks have been proposed \cite{Resta99_Electron_localization, Gurarie11_Interacting_TIs, man12, Pollmann10_Entaglement_spectrum_of_topological_phases_in_1D, Wen12_Topological_classification_of_non_interacting_fermions}.  For this reason, the analysis of simple models where exact solutions can be obtained is of utmost importance for the development of theory.
\begin{figure}[t]
\centering

\begin{subfigure}{\columnwidth}
\caption{\hspace{80mm}\color{white}.\color{black}} 
\begin{tikzpicture}
    \node at (0.2,2.0) {\includegraphics[scale=0.38]{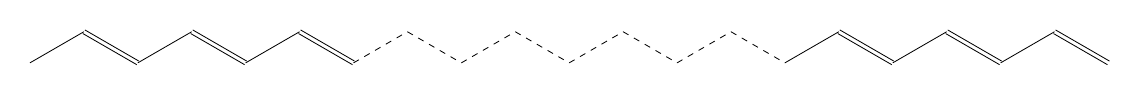}};
    \draw[C0, fill=C0!20,thick] (-1.3,1.5) rectangle (1.6,1.7); 
    \draw[-,thick,C0] (0,1.5)--(0,1);
    \draw[-,thick,C0] (-0.5,1)--(0.5,1);
    \draw[-,thick,C0] (-0.3,0.9)--(0.3,0.9);
    \draw[-,thick,C0] (0,0.9)--(0,0.5);
    \draw[-,thick,C0] (-0.3,0.5)--(0.3,0.5);
    \draw[-,thick,C0] (-0.2,0.4)--(0.2,0.4);
    \draw[-,thick,C0] (-0.1,0.3)--(0.1,0.3);
    \node at (-0.8,1) {$V_g$};
\end{tikzpicture}
\label{fig:subfig2}
\end{subfigure}
\hfill

\begin{subfigure}{\columnwidth}
\caption{\hspace{80mm}\color{white}.\color{black}} 
\begin{tikzpicture}
    \draw[->] (0,-3)--(0,1) node[above]{$V(x)$};
    \draw[->] (-3,0)--(3,0) node[right]{$x$};
    \draw[-,line width=2 pt,C1] (-2.8,0)--(-1.5,0);
    \draw[-,line width=2 pt,C1] (-1.5,0)--(-1.5,-2);
    \draw[-,line width=2 pt,C1] (-1.5,-2)--(1.5,-2);
    \draw[-,line width=2 pt,C1] (1.5,-2)--(1.5,0);
    \draw[-,line width=2 pt,C1] (1.5,0)--(2.8,0);
    \node [below] at (-0.5,-2.3) {$-V_g$};
    \node[above] at (-1.5,0) {$-d/2$};
    \node[above] at (1.5,0) {$d/2$};
\end{tikzpicture}
\label{fig:subfig1}
\end{subfigure}

\caption{(Color online) (a) Schematic representation of the device. A linear tPA molecule capacitively coupled to a potential gate. The dotted lines represent the formation of DWs. (b) Electrostatic potential well of width $d$ and depth $V_g$ induced by the gate.} 
\label{fig:device}
\end{figure}
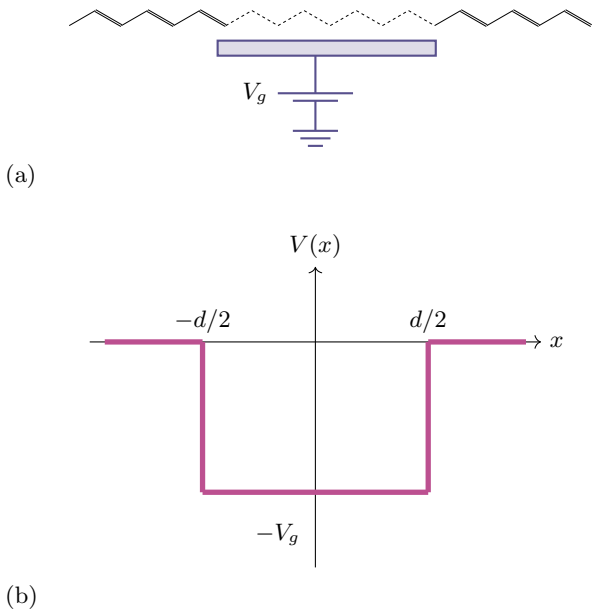

Motivated by these fundamental questions as well as by the recent experimental advances in the field of organic-electronic devices, in this work we study an interacting tPA chain coupled to an external gate voltage (see Fig. \ref{fig:device}) and analyze its topological excitations and the topological quantum phase diagram. To that end, we describe the system by the means of the Takayama-Lin-Liu-Maki (TLM) Hamiltonian \cite{Takayama1980a}, a continuum version of the SSH Hamiltonian, and analyze it in the framework of the Abelian bosonization formalism \cite{Giamarchi2004, Gogolin1998}, a powerful field-theoretical technique allowing us to obtain an analytical description of strongly-interacting one-dimensional quantum phases. We obtain the ground state of the device by solving a modified sine-Gordon equation depending on the parameter $V_g$ and the width of the gate $d$, for both the charge- and spin-density bosonic fields [$\phi_c(x)$ and $\phi_s(x)$, respectively] obtained from their classical Euler-Lagrange equations of motion. While in the absence of a gate the topological class of the SSH model is usually described by a $\mathbb{Z}_2$ topological number \cite{Asboth16_Short_course_on_TIs}, the introduction of a gate potential that locally breaks particle-hole symmetry changes the classification to a $\mathbb{Z}$ topological invariant $q$, whose value simultaneously indicates both the excess charge and the number of DWs induced at the gate and can be controlled by $V_g$  and the geometrical parameters of the device \cite{Arancibia25_Towards_electrical_DW_control}.

The rest of the paper is organized as follows: In Sec. \ref{sec:theoretical_framework}, we introduce the theoretical model for a tPA molecule capacitively coupled to an external gate voltage and introduce the bosonization procedure.  In Sec. \ref{sec:eqs_motion} we derive the  classical equations of motion of the system and obtain their solutions (a series of multikink states with different number of DWs) representing  the ground- and excited-state configurations. In Sec. \ref{sec:topology} we consider the definition of a proper topological invariant (i.e., a winding number $\mathbb{Z}$) in bosonized form, allowing the topological classification of the ground states of our system and discuss the crucial effect of the local particle-hole symmetry breaking induced by the gate. In Sec. \ref{sec:results} we summarize our results and analize the topological character of the solutions, the energy of the ground state and its topological charge, and the topological phase diagram in terms of the experimentally tunable parameters of the system and the Coulomb interaction strength. Finally, in Sec. \ref{sec:summary}, we provide a summary of our conclusions and some perspectives.

\section{Theoretical framework}\label{sec:theoretical_framework}

We focus on the system depicted in Fig. \ref{fig:device}, consisting of a single tPA molecule capacitively coupled to an external gate voltage of width $d$. We model this system using the TLM Hamiltonian in the continuum \cite{Takayama1980a}, modified to include both the effect of repulsive short-range (i.e., contact) interactions  among electrons and the effect of the gate:
\begin{equation}
  H= H_{e} + H_\text{latt}+H_{e\text{-latt}}+H_{e\text{-g}} \label{eq:H}.
\end{equation}
The Hamiltonian describing the energy of the classical lattice configuration in Eq. (\ref{eq:H}) is \cite{Takayama1980a}
\begin{align}
H_\text{latt}=\kappa \int_{-L/2}^{L/2} dx\, \Delta^2(x),\label{eq:H_latt}
\end{align}
where $\Delta(x)$ is a classical real scalar field representing the continuum version of the staggered dimerization amplitude of the ions in the discrete SSH chain \cite{Su79_Solitons_in_polyacetylene}, i.e., $\Delta(x)\simeq (-1)^n \alpha_{e\text{-latt}} (u_{n+1}-u_n)$, where $\alpha_{e\text{-latt}}$ and $u_n$ are defined in the microscopic SSH model and correspond to the electron-lattice interaction and to the displacement of the $n$th CH group from equilibrium, respectively. In other words, the value and magnitude of $\Delta(x)$ gives information about the dimerization state of the lattice at the point $x$. The parameter $\kappa$ represents the elastic coefficient in the continuum model  \cite{Takayama1980a}. Since this field describes the heavier (as compared with electrons) --CH groups in the tPA molecule, we have assumed the Born-Oppenheimer approximation and the dynamics of $\Delta(x)$ has been neglected \cite{Takayama1980a,Heeger1988a}.

The term $H_{e}=H_0+H_\text{int}$ describes one-dimensional interacting electrons represented by the fermionic annihilation field $\psi_\sigma(x)$ with $\sigma=\{\uparrow, \downarrow \}$ the spin projection. Here $H_0$ represents the electronic kinetic energy obtained from the usual linearization of the band structure around the Fermi points $\pm k_F$ \cite{Giamarchi2004, Gogolin1998}. 
The linearization procedure enables a low-energy description of the field $\psi_\sigma(x)$ in terms of right- and left-moving fields $\psi_{\sigma \alpha}(x)$  ($\alpha=\{ R,L\}$) with spin $\sigma\in\{\uparrow,\downarrow\}$, i.e., $\psi_\sigma(x)=e^{ik_Fx}\psi_{\sigma R}(x) + e^{-ik_Fx}\psi_{\sigma L}(x)$. The explicit form of the electron-electron interaction term $H_\text{int}$, including forward-, umklapp- and back-scattering interactions, is given in Appendix \ref{app:fermions}.

The term $H_{e\text{-latt}}$ describes the electron-lattice interaction, of crucial importance here. The sum of the terms $H_0+H_{e\text{-latt}}$ corresponds to two Kramers-degenerate copies  of the massive Dirac model in 1 spatial dimension, where the field $\Delta(x)$ effectively represents a position-dependent mass term, a paradigmatic model studied by Jackiw and Rebbi \cite{Jackiw76_Jackiw_Rebbi_soliton}:
\begin{multline}
H_0+H_{e\text{-latt}} = \\ \sum_\sigma \int_{-L/2}^{L/2} dx \,\Psi_\sigma^\dagger(x)\left[ -i v_\mathrm{F}\tau_z\partial_x +\tau_x \Delta(x)\right] \Psi_\sigma(x) 
\end{multline}
where $\tau_i$ are the Pauli matrices and $\Psi_\sigma^\dagger = ( \psi_{\sigma R}^\dagger, \psi_{\sigma L}^\dagger )$ is the spinor representation of the fermionic field.

The presence of a gate voltage, capacitively coupled to the tPA molecule, is represented by the term 
\begin{equation}
    H_{e\text{-g}}=\int_{-L/2}^{L/2} dx\, V(x)\rho(x),\label{eq:H_eg}
\end{equation}
where $\rho(x)=\sum_{\sigma\alpha} \rho_{\sigma\alpha}(x)$ is the total electronic density $\rho_{\sigma\alpha}=\psi^\dagger_{\sigma\alpha}\psi_{\sigma\alpha}$ and $V(x)$ is the potential induced by the gate voltage, which for simplicity we have assumed to take a potential-well form [see Fig. \ref{fig:device}(b)]:
\begin{equation}\label{eq:gate}
    V(x) = -V_g\left[\Theta_{H}(x+d/2)-\Theta_{H}(x-d/2)\right],
\end{equation}
where $V_g$ is the depth of the well, $d$ its width, and $\Theta_H(x)$ the Heaviside step-function. While more realistic models to describe the tPA-gate capacitive interaction could be introduced [e.g., describing a softer potential $V(x)$], our main qualitative conclusions will not be modified due to the topological character of the solutions (see Sec. \ref{sec:summary}). 

Physically, the introduction of a gate will have a non-trivial effect on the system. Intuitively, the increase of $V_g>0$ (i.e., the increase of the potential well depth $-V_g$) should favor the accumulation of electric charge in the gated region of the molecule. However, since the system is a Peierls insulator, this charge accumulation competes with the  (dimerization) gap, which does not allow the charge to change continuously due to the absence of a continuum of states near the Fermi energy. In addition, the gate will affect indirectly the lattice degrees of freedom via the electron-lattice coupling term in Eq. (\ref{eq:H}), which introduces nontrivial effects on the ground-state configuration. In the next section we analyze this effect on the quantum phase diagram in detail.

Within this continuum model, the essential physics of the original SSH model can be described by the sum of the terms $H_0+H_{e\text{-latt}} +H_\text{latt}$, that is, Eq. (\ref{eq:H}) in the absence of interactions and in the absence of the gate potential \cite{Su79_Solitons_in_polyacetylene, Su80_Soliton_excitations_in_PA, Takayama1980a}.  In the case of a uniform distortion that minimizes the total ground-state energy, $\Delta(x)\equiv \Delta_0$, the system can be shown to be doubly-degenerate under the change $\Delta_0 \to -\Delta_0$ and the simultaneous unitary transformation $\psi_{\sigma R}\to e^{-i\pi/2}\psi_{\sigma R}$, $\psi_{\sigma L}\to e^{i\pi/2}\psi_{\sigma L}$, corresponding to the two possible Peierls-dimerized ground states of the system. A soliton excitation in this context corresponds to a configuration of $\Delta(x)$ that smoothly connects the uniform configuration $- \Delta_0$ in one portion of the system to the configuration $\Delta_0$ at another portion, and the point $x_0$ at which  $\Delta(x_0)=0$ corresponds to the center of a DW. Standard theoretical treatments of either the SSH or the TLM models typically assume a specific configuration of the lattice, corresponding to a frozen configuration of the dimerization field $\Delta(x)$, either $\Delta_0$ or $-\Delta_0$ (see, i.e., Ref. \cite{Asboth16_Short_course_on_TIs}). However, we must keep in mind that the actual configuration of $\Delta(x)$ in the ground state must be obtained from the minimization of the total energy of the system, encompassing both electronic and lattice degrees of freedom. A distinctive aspect in this work is the fact that the field $\Delta(x)$ itself gets modified in response to the applied gate potential $V_g$, in order to minimize the global energy in the ground state. Since the presence of the gate  breaks both translation and particle-hole symmetry, finding the ground-state configuration of $\Delta(x)$  is  a crucial nontrivial step (see Sec. \ref{sec:eqs_motion}). This is a key difference with respect to other approaches, with important consequences for the topological properties of the device, as we will see below.

Finally, we mention that although the physically relevant problem we aim to study involves a finite tPA molecule with open boundaries, the bosonization formalism becomes technically cumbersome in that case \cite{Dolcini05_Transport_in_wires_with_impurity, Lecheminant02_Finite_spin_ladders} and to simplify our calculations we assume closed boundary conditions. Since the gate voltage is applied away from the edges, we can reasonably assume that the large-$L$ limit reproduces the correct physical behavior irrespective of the chosen boundary conditions. An important caveat, however, is that since the sign of the field $\Delta(x)$ must change at each DW,  to reproduce a physical situation with an even number of DWs induced by the gate in the ground state, the only physically consistent boundary condition is the periodic boundary condition, i.e., $\Delta(-L/2)=\Delta(L/2)$. In contrast, to describe configurations where the gate induces an odd number of DWs, antiperiodic boundary conditions $\Delta(-L/2)=-\Delta(L/2)$ are necessary. This is, however, a minor price to pay for the substantial technical simplification gained. Regarding the electronic degrees of freedom, in general twisted boundary conditions $\psi_{\sigma \alpha}(x+L)=e^{i 2 \pi n_{\sigma \alpha}}\psi_{\sigma \alpha}(x)$, with $n_{\sigma \alpha}$ a fractional number, must be imposed. We return to this point in Sec. \ref{sec:eqs_motion}.

To investigate the low-energy properties of the system, we employ the standard Abelian bosonization formalism (see Appendix \ref{app:fermions} for details). 
In this framework, the low-lying excitations are described by a pair of bosonic phase fields $\phi_\nu(x)$, where $\nu = c, s$ denotes the charge and spin 
collective modes, respectively. A standard manipulation after bosonization renders the electronic Hamiltonian in the form $H_e = H_s + H_c$ \cite{Giamarchi2004, Gogolin1998}, which explicitly decouples the charge and spin modes, with each $H_\nu$ taking the form of a standard sine-Gordon model:
\begin{multline}
   H_\nu=\frac{v_\nu}{2\pi}\int_{-L/2}^{L/2} dx\,\left[\frac{1}{K_\nu}(\partial_x\phi_\nu(x))^2+K_\nu\pi^2\left(\Pi_\nu(x)\right)^2\right]\\
   +\frac{2 g_\nu}{(2\pi a)^2}\int_{-L/2}^{L/2} dx\,\cos \sqrt{8}\phi_\nu(x).
   \label{eq:H_e}
\end{multline}
In this expression, $\Pi_\nu(x)$ is the canonically conjugate momentum to $\phi_\nu(x)$, $v_\nu$ physically represents the velocity of charge or spin collective modes and $K_\nu$ are the dimensionless ``Luttinger parameters'', that encode forward-scattering interactions. The couplings $g_s$ and $g_c$ denote backscattering and umklapp interactions respectively \footnote{The usual notation in the literature is $g_s=g_1$ and $g_c=g_3$. We have chosen a different, more compact notation.}. In terms of phase fields $\phi_\nu(x)$ the electron charge- and spin- density operator is written as $\rho_\nu(x)=-\sqrt{2}\partial_x\phi_\nu(x)/\pi$.

 The electron-lattice interaction and the coupling to the external potential within this framework are respectively given by
\begin{align}
    H_{e\text{-latt}}&=\frac{2}{\pi a}\int_{-L/2}^{L/2} dx\,\Delta(x)\cos \sqrt{2}\phi_c(x)\cos\sqrt{2}\phi_s(x),
    \label{eq:H-e-latt}
\end{align}
and
\begin{align}
    H_{e\text{-g}}&=-\frac{\sqrt{2}}{\pi}\int_{-L/2}^{L/2} dx\,V(x)\partial_{x}\phi_c(x).
    \label{eq: H_el_g}
\end{align}
In this last term we have made the experimentally reasonable assumption that $d\gg k_F^{-1}$, which allows to safely neglect rapidly oscillating terms $\sim e^{i2k_Fx}$ in the charge density. Note that from  perturbative (i.e., weak-coupling) scaling analysis, the scaling dimensions of the backscattering and umklapp terms are higher than that of the electron-lattice term Eq. (\ref{eq:H-e-latt}), meaning that this is the dominant interaction at low energies \cite{Sachdev_book}.  However, backscattering and umklapp can induce non-negligible renormalizations of the bare electron-lattice coupling parameters, as we demonstrate in the next section.

\section{Equations of motion and solutions in the presence of a gate}\label{sec:eqs_motion}

We now focus on the derivation of the interacting  ground state of the device. We note that the 1D Hamiltonian couples only the commuting fields $\phi_s(x)$ and $\phi_c(x)$ which, in the static case,  allows a simultaneous  minimization of the total energy of the system. In other words, the Hamiltonian has a well-defined classical limit given in terms of the fields $\phi_s(x)$ and $\phi_c(x)$, from which valuable insights into the strongly-correlated ground state can be obtained. The Hamilton equations of motion $\dot{\phi}_\nu(x) = \frac{\delta H}{\delta \Pi_\nu(x)} $, $\dot{\Pi}_\nu(x)= -\frac{\delta H}{\delta \phi_\nu(x)}$, can be expressed  in the static case  as
\begin{multline}
\frac{v_\nu}{\sqrt{2}K_\nu}\partial_x^2\phi_\nu(x) + \frac{2}{a}\Delta(x)\sin\sqrt{2}\phi_\nu(x)\cos\sqrt{2}\phi_{\bar{\nu}} (x) \\ 
    + \frac{g_\nu}{\pi a^2}\sin\sqrt{8}\phi_\nu(x)
    = \delta_{\nu c}\partial_x V(x).\label{eq:EOMs}
\end{multline}
An analog procedure applied to the field $\Delta(x)$ reduces, under the simplifying assumptions of a static field, to the energy-minimum condition $\frac{\delta H}{\delta \Delta(x)} = 0$, which yields the equation:
\begin{align}
    \Delta(x) &= -\Delta_0\cos\sqrt{2}\phi_c(x)\cos\sqrt{2}\phi_s(x),
    \label{eq: EOM delta}    
\end{align}
where we have defined the parameter $\Delta_0=8 \alpha_{e\text{-latt}}  u_0=1/(\kappa \pi a)$. Together, Eqs. (\ref{eq:EOMs}) and(\ref{eq: EOM delta}) can be considered as a self-consistent system of equations where the configuration of the lattice degrees of freedom determine the electronic ones, and vice versa. Interestingly, in the framework of Abelian bosonization, we can simply substitute Eq. (\ref{eq: EOM delta}) into Eq. (\ref{eq:EOMs}), and reduce the system of coupled equations to:
\begin{multline}
    \frac{v_\nu}{\sqrt{2}K_\nu}\partial_x^2\phi_\nu(x) - \sin\sqrt{8}\phi_\nu(x)\left(\frac{\Delta_0}{a}\cos^2\sqrt{2}\phi_{\bar{\nu}}(x)  
    - \frac{g_\nu}{\pi a^2}\right)\\
    = -\delta_{\nu c} V_g \left[\delta(x+d/2)-\delta(x-d/2)\right],\label{eq:EOMsDecoupled}
\end{multline}
where we have used the explicit form of the potential in Eq. (\ref{eq:gate}) and introduced the notation $\bar{c}=s$ and $\bar{s}=c$. Equation (\ref{eq:EOMsDecoupled}) determine the ground-state configurations of the fields $\phi_\nu(x)$, which are defined modulo $\pi/\sqrt{2}$. 
The effect of the gate voltage from Eq. (\ref{eq:gate}) is entirely captured by
the right-hand term in Eq. (\ref{eq:EOMsDecoupled}), where the  Dirac delta functions induce discontinuities in the first derivatives of the field $\phi_c(x)$ at the edges of the potential well. Then, the ground-state energy of the system can be computed by substituting the static saddle-point solutions into the classical limit of the Hamiltonian, obtained by setting all terms involving the canonically conjugate momentum terms $\Pi_\nu(x) \to 0$ in the full quantum Hamiltonian $H$ in Eq. (\ref{eq:H}). Finally, we substract the constant contribution $E_0 = \left(-\frac{\Delta_0}{\pi a} + \frac{g_c+g_s}{2(\pi a)^2}\right) L$, formally divergent in the limit $L\rightarrow \infty$, arising from the uniform configurations of the fields $\phi_\nu(x)=\phi_\nu^{(0)}(x)$ [see Eqs. (\ref{eq:uniform})] which correspond to the ground state of the system in the absence of gate voltage. In this way, the ground state energy is defined as the energy \textit{on top} of $E_0$, and therefore is well-defined for the localized topological excitations we consider below. 

Finally, the total number of particles and the total spin, measured with respect to the uniform background electronic density of the half-filled Fermi sea, i.e.,  $\rho_{0,c}=a^{-1}$, $\rho_{0,s}=0$, are given by the integral of the densities $\rho_\nu(x)$, i.e.,
\begin{align}
   N_\nu &= \int_{-L/2}^{L/2} dx\,\rho_\nu(x) = - \frac{\sqrt{2}}{\pi} \int_{-L/2}^{L/2} dx\,\partial_x\phi_\nu(x) \nonumber \\ & = - \frac{\sqrt{2}}{\pi}\left[\phi_\nu\left(\frac{L}{2}\right)-\phi_\nu\left(-\frac{L}{2}\right)\right].
   \label{eq:Ns}
\end{align}
However, note that these quantities cannot take arbitrary values because they are constrained by the boundary conditions imposed on the field $\Delta(x)$. As mentioned Sec. \ref{sec:theoretical_framework}, the dimerization field $\Delta(x)$ must change sign across each DW. Assuming that the gate-induced DWs are far from the edges of the system, such property is ensured by imposing the boundary conditions $\Delta(-L/2)=\pm \Delta(L/2)$. According to Eq. (\ref{eq: EOM delta}), the above conditions imply the boundary condition on the bosonic fields $\phi_{\nu}(x+L)=\phi_{\nu}(x)+\pi N_{\nu}/\sqrt{2}$, with $N_\nu$ an integer number. Then, we see that the total excess charge $Q=-e N_c$ becomes quantized in units of $e$, and the total spin $S_z=(1/2) N_s$ becomes quantized in units of 1/2 (here we have assumed $\hbar=1$). Finally, making use of the bosonization identities in Eqs.~(\ref{eq:bosonization_formula}) 
and (\ref{eq:field_transformation}) (see Appendix~\ref{app:fermions}), the fermionic phase shifts arising from the boundary conditions $\psi_{\sigma \alpha}(x+L)=e^{i 2 \pi n_{\sigma \alpha}}\psi_{\sigma \alpha}(x)$
are related to $N_\nu$ through 
\begin{align}
    N_c&=n_{\uparrow,R}+n_{\downarrow,R}+n_{\uparrow,L}+n_{\downarrow,L},\\
    N_s&=n_{\uparrow,R}-n_{\downarrow,R}+n_{\uparrow,L}-n_{\downarrow,L}.
\end{align}
From here,  it is relatively easy to see that in the presence
of SU(2) and inversion symmetries, the quantities $n_{\sigma, \alpha}$ must be equal to $n_{\sigma, \alpha}=N_c/4$, indicating that fermions $\psi_{\sigma \alpha}(x)$ obey twisted boundary conditions.

\subsection{Absence of gate potential: Recovering known results for tPA}

Before analyzing the effects of the gate voltage, it is instructive to consider the case of $V_g=0$, where the system recovers translational and particle-hole symmetry. In this case, it is straightforward to see from Eq. (\ref{eq:EOMsDecoupled})  that the uniform configurations given by 
$(\phi_c^{(0)}(x),\phi_s^{(0)}(x)) = \left(\frac{p_c\pi}{\sqrt{8}},\frac{p_s\pi}{\sqrt{8}}\right)$, where $p_c,p_s \in \mathbb{Z}$, are solutions of the equations of motion. In particular, to minimize the energy, $p_c,p_s$ must be even integers, so 
\begin{align}
(\phi_c^{(0)}(x),\phi_s^{(0)}(x)) &= \left(\frac{m_c\pi}{\sqrt{2}},\frac{m_s\pi}{\sqrt{2}}\right),\label{eq:uniform}
\end{align}
with $m_c, m_s \in \mathbb{Z}$. 
Note that these uniform configurations obey the periodic condition $\phi_{c(s)}(L/2)-\phi_{c(s)}(-L/2)=0$, implying from Eq. (\ref{eq:Ns}) that we must have $N_c=N_s=0$ and, therefore the total charge and spin in the ground state is $Q=S_z=0$, as expected for a particle-hole- and SU(2)-symmetric ground state.

Substituing the uniform solution into Eq. (\ref{eq: EOM delta}), the dimerization pattern becomes $\Delta(x) = - \Delta_0 (-1)^{m_c+m_s}=\pm \Delta_0$. Due to our definition, the energy of these configurations is zero and independent of $m_c,m_s$ and therefore independent of the sign of the dimerization parameter, reflecting the doubly-degenerate nature of the uniformly Peierls-dimerized groundstate of tPA mentioned earlier. 

Beyond the uniform solutions described above, the general analytical form of the solutions to the coupled sine-Gordon-type equations (\ref{eq:EOMsDecoupled}) is very complicated. However,  in the weakly-interacting case where $\Delta_0 \gg \{g_{\sigma \sigma^\prime, \alpha \alpha^\prime}/a, g_c/a, g_s/a\}$ [where $g_{\sigma \sigma^\prime, \alpha \alpha^\prime}$ are the forward-scattering interactions, see Eq. (\ref{eq:geology})] a reasonable assumption leading to the lowest-lying excitations corresponds to freezing either $\phi_c^{(0)}(x)$ or $\phi_s^{(0)}(x)$  in the uniform configuration, while the other one satisfies the sine-Gordon equation
\begin{align}
 \partial_x^2\phi_\nu(x) -\frac{1}{\sqrt{8}\xi^2_\nu}\sin\sqrt{8}\phi_\nu(x)&= 0, \quad (\nu=s,c),\label{eq:sine_gordon}    
\end{align}
where we have defined the localization lengths
\begin{align}
\xi^{-2}_\nu&=\frac{4K_\nu}{v_\nu}\left(\frac{\Delta_0}{a}-\frac{g_\nu}{\pi a^2}\right)
\end{align}

When either the charge or spin sector is frozen in the trivial uniform solution, the simplest non-trivial excitation for the remaining degree of freedom in infinite space given by Eq. (\ref{eq:sine_gordon}) corresponds to a solitonic configuration centered at position $x_0$:
\begin{equation}
    \phi_\nu(x) = \pm \sqrt{2} \arctan\left[\exp\left(\frac{x-x_0}{\xi_\nu}\right)\right]+\frac{m_\nu\pi}{\sqrt{2}} \label{eq: soliton},
\end{equation}
where the width is characterized by $\xi_c$. The $+$ ($-$) sign corresponds to a soliton (antisoliton) solution which interpolates between the uniform configurations  $\phi_\nu(-\infty)=\frac{m_\nu\pi}{\sqrt{2}}$ and $\phi_\nu(+\infty)=\frac{(m_\nu\pm 1)\pi}{\sqrt{2}}$, when we take the limit $L\rightarrow \infty$.  Note that due to the translational invariance in the absence of gate voltage, there is a continuum of degenerate solutions defined by the value of $x_0$. This excitation has a classical energy 
\begin{align}
    E_\text{sol}^\nu&=2\sqrt{\frac{v_\nu}{\pi K_\nu}\left[\frac{\Delta_0}{\pi a}-\frac{g_\nu}{(\pi a)^2}\right]}=\frac{v_\nu}{\pi K_\nu \xi_\nu}.\label{eq:E_soliton}
\end{align}
Due to our assumption that $\Delta_0 \gg \{g_s/a,g_c/a\}$, note that the energy of the soliton excitation is finite.

In the specific case where $v_s/K_s=v_c/K_c$ and $g_c=g_s$, Eqs. (\ref{eq:EOMsDecoupled}) are symmetrical under the exchange $\phi_\nu(x) \leftrightarrow \phi_{\bar{\nu}}(x)$, and therefore become degenerate in energy. An important particular case of this situation is the non-interacting case where
$K_c=K_s=1$, $v_c=v_s=v_F$, and $g_c=g_s=0$. This specific case has already been discussed in Refs. \cite{Su80_Soliton_excitations_in_PA, Heeger1988a}, and correspond to the well-known soliton excitations with fractionalized quantum numbers with either charge $Q=\pm e$ and spin $S=0$, or $Q=0$ and spin $S=\pm 1/2$. 
These results are precisely recovered when the solitonic solutions (\ref{eq: soliton}) are replaced into Eq. (\ref{eq:Ns}).

\subsection{Effect of the gate voltage: Derivation of the modified sine-Gordon equation}

We now consider the case of an applied gate. Since the  gate voltage $V_g$ only couples to the charge field through the density field $\partial_x \phi_c(x)$ in Eq. (\ref{eq:EOMsDecoupled}) and cannot induce spin excitations, in what follows we assume that the field $\phi_s(x)$ remains in the trivial uniform configuration. In this case, Eq. (\ref{eq:EOMsDecoupled}) for the charge field simplifies to
\begin{multline}
 \partial_x^2\phi_c(x) -\frac{\sin\sqrt{8}\phi_c(x)}{\sqrt{8}\xi^2_c} \\
 = -\frac{1}{\xi_g} \left[\delta\left(x+\frac{d}{2}\right)-\delta\left(x-\frac{d}{2}\right)\right],\label{eq:EOM phic final} 
\end{multline} 
where the effect of the gate voltage is totally encoded in the new lengthscale
\begin{equation}
    \xi_g \equiv \frac{v_c}{\sqrt{2}K_c}\frac{1}{V_g}.\label{eq:xi_g}
\end{equation}
Defining a dimensionless coordinate $\tilde{x}=x/\xi_c$, Eq. (\ref{eq:EOM phic final}) becomes
\begin{multline}
\partial_x^2\phi_c\left(\tilde{x}\right) -\frac{\sin\sqrt{8}\phi_c\left(\tilde{x}\right)}{\sqrt{8}} \\
= -\frac{\xi_c}{\xi_g} \left[\delta\left(\tilde{x}+\frac{d}{2\xi_c }\right)-\delta\left(\tilde{x}-\frac{d}{2\xi_c }\right)\right],\label{eq:EOM_phic_final_dimensionless} 
\end{multline} 
i.e., the energy of the ground state only depends on the ratios of lengthscales $\xi_c/\xi_g$ and $d/\xi_c$. These ratios physically encode a nontrivial interplay among competing mechanisms, such as gating effects, Coulomb repulsion, and Peierls dimerization.

Note that the trivial ground states (\ref{eq:uniform}) are no longer valid solutions in the presence of an applied gate voltage, and spatially varying configurations must be considered. Since the gate potential $V(x)$ is localized in the region $(-d/2,d/2)$, in the limit  $L\rightarrow \infty$  the field $\phi_c(x)$ can be assumed to be unaffected in regions far from the gate. In that case we require $\phi_c(x)$ to interpolate between trivial vacuum states given by Eq. (\ref{eq:uniform}) and to obey the boundary conditions:
\begin{align}
\partial_{x}\phi_c\left(x=\pm \frac{L}{2}\right)&= 0,\label{eq:boundary_conditions_infinity1}\\ 
\cos\left[\sqrt{8}\phi_c\left(x=\pm \frac{L}{2}\right)\right]&=1.
\label{eq:boundary_conditions_infinity2}
\end{align}
The first condition ensures that the solution is localized and has a finite energy, whereas the second imposes that the dimerization field $\Delta(x)$ only takes the values $\pm\Delta_0$ far from the gate, thus recovering the physics of the Peierls dimerization. In addition, due to the discontinuity introduced by the gate voltage $V_g$ at points $x=\pm d/2$, we need to find the appropriate boundary conditions at the interface between gated and nongated regions. Integrating Eq. (\ref{eq:EOM phic final}) around the discontinuity points we find:
\begin{align}
    \partial_x\phi_c(x=\pm d/ 2^+)-\partial_x\phi_c(x=\pm d/2^-) &= \pm \xi_g^{-1},
    \label{eq: gluing_derivative}\\
    \phi_c(x=\pm d/ 2^+)-\phi_c(x=\pm d/2^-)&=0,\label{eq: gluing_function}
\end{align}
where the last condition (i.e., continuity at the interface) is obtained by integrating the first one.

\subsection{General solution in the presence of the gate}

To simplify the derivation of the general solution of  Eq. (\ref{eq:EOM phic final}), we first note that the potential $V(x)$ in Eq. (\ref{eq:gate}) is even under the inversion $x \rightarrow -x$. Then, the energy of the ground state is minimized when the charge density is an even function, i.e.,  $\partial_x \phi_c(x)=\partial_x \phi_c(-x)$ [see Eq. (\ref{eq: H_el_g})]. Integrating this equation, we obtain that $\phi_c(x) = -\phi_c(-x) + C$, 
where $C$ is an integration constant. This generic form ensures that the Hamiltonians $H_{0,\nu}$ and $H_\text{e-g}$ [Eqs. (\ref{eq:Hnu}) and (\ref{eq: H_el_g}), respectively] remain invariant under space inversion. However, due to the presence of the $\cos \sqrt{8}\phi_c(x)$ inside the terms $H_e^{\text{um}}$ and $H_{e-\text{latt}}$, the constant $C$ cannot be chosen arbitrarily and must take the form  $C = \frac{q \pi}{\sqrt{2}}$, with $q\in\mathbb{Z}$, for these terms to remain invariant as well. From these symmetry requirements, we arrive at the general form obeyed by the field $\phi_c(x)$ in the ground state:
\begin{align}
\phi_c(x) &= -\phi_c(-x) + \frac{q\pi}{\sqrt{2}}. \label{eq:symmetry_constraint}
\end{align}
Evaluating this expression for $x=L/2$, we obtain the equation $\phi_c(L/2)+\phi_c(-L/2)=q \pi/\sqrt{2}$, which along with Eq. (\ref{eq:Ns}), gives
\begin{align}
\phi_c(L/2)&=\frac{\pi}{\sqrt{2}}\frac{q-N_c}{2},\label{eq:phi_c_L_2}\\
\phi_c(-L/2)&=\frac{\pi}{\sqrt{2}}\frac{q+N_c}{2}.\label{eq:phi_c_minus_L_2}
\end{align}
Since the field $\phi_c(x)$ is globally defined only modulo $\pi/\sqrt{2}$, we can, without loss of generality, fix its value far from the gate to one of the vacuum solutions $m_c \pi/\sqrt{2}$ in Eq. (\ref{eq:uniform}), where the solution is constant. In particular, we choose the value $\phi_c(L/2)=0$, which corresponds to $m_c=0$. This choice is convenient since from Eq. (\ref{eq:phi_c_L_2}) we obtain the relation $q=N_c$, and since the integer $q$ is a property of the gate via Eq. (\ref{eq:symmetry_constraint}), we conclude that the gate is reponsible for changing the total number of the electrons in the system. This important property allows to interpret the integer $q$ as a $\mathbb{Z}$ topological invariant which enables the topological classification of our solutions (see Sec. \ref{sec:summary} for a more detailed discussion).

To construct the general solution compatible with the symmetries of the system, we divide the space into three regions according to the form of the potential $V(x)$, i.e.,  $R_1 = (-L/2,-d/2)$, $R_2 =(-d/2,d/2)$,  and $R_3 = (d/2,L/2)$, and look for the most general solution in each region. Since we are interested in the limit $L\rightarrow \infty$, in regions $R_1$ and $R_3$,  we can use the soliton solutions of Eq. (\ref{eq: soliton}). However, a different family of solutions must be considered in region $R_2$ in order to satisfy the boundary conditions at the edges of the gate. To that end, we note that besides the usual kink,  the sine-Gordon equation (\ref{eq:sine_gordon}) admits more general solutions that are periodic with period $\vert X_k\vert$ modulo $\pi/\sqrt{2}$ \cite{Manton2010}, that is,  
\begin{align} \phi_c(x)&=\phi_c(x - X_k)+\frac{\pi}{\sqrt{2}}.
\end{align}
Then, the most general solution in $R_2$  is 
\begin{align}
    \phi_c(x)&=-\frac{1}{\sqrt{2}}\am\left[\frac{ k (x-x_0)}{\xi_c}  \Bigg\vert -k^{-2}\right].
\end{align}
where the Jacobi amplitude function $\am(u|m)$ is related to the incomplete elliptic integral of the first kind $F(u,m)$ through the relation $F\bm{(}\am(u|m),m\bm{)}=u$, and $X_k=2\xi_c K(-k^{-2})/k$, with $K(m)$ the complete elliptic integral of the first kind \cite{abramowitz} (see also Appendix \ref{app:Jacobi} for details). Note that $X_k$ is negative when $k<0$; in this case we obtain a kink-like solution. For $k>0$ we obtain an  antikink-like solution. The parameter $x_0$ is an arbitrary constant which can be found through the boundary conditions Eqs. (\ref{eq: gluing_derivative}) and (\ref{eq: gluing_function}). As we show below, the multikink solutions of the modified sine-Gordon Eq. (\ref{eq:EOM phic final}) correspond to multi-DW configurations, which are energetically forbidden in the absence of the gate \cite{Rajaraman1984}.

After applying the boundary conditions Eqs. (\ref{eq:boundary_conditions_infinity1}) and (\ref{eq:boundary_conditions_infinity2}) at $x=\pm L/2$  in regions $R_1$ and $R_3$, and enforcing the  symmetry constraint given by Eq. (\ref{eq:symmetry_constraint}), the most general solution can therefore be expressed as
\begin{align}
    \phi_c(x) &= \begin{cases}
         \eta\sqrt{2} \arctan\left[\exp \left(\frac{x+b}{\xi_c}\right)\right] + \frac{q\pi}{\sqrt{2}}, & x\in R_1, \\
         \frac{-1}{\sqrt{2}}\am \left[\frac{k}{\xi_c}\left(x- q\frac{ X_k}{2}\right)\big|-k^{-2}\right],  & x\in R_2, \\
         -\eta\sqrt{2} \arctan\left[\exp \left(-\frac{x-b}{\xi_c}\right)\right],  & x\in R_3,
    \end{cases}\label{eq:generic_solution}
\end{align}
where  $\eta$ is a parameter that specifies either a soliton-like ($\eta = +1$) or an antisoliton-like ($\eta = -1$) solution, and  $b$ is a parameter that determines its localization center.  The parameter $k$ is also an integration constant, to be determined imposing the boundary conditions. Finally, 
we note that the case of a constant configuration in regions $R_1$ and $R_3$ (obtained by taking $\eta = 0$), although possible, is only compatible with the boundary conditions, Eqs. (\ref{eq:boundary_conditions_infinity1}) and (\ref{eq:boundary_conditions_infinity2}) at specific values of $V_g$.

We now impose the boundary conditions at the interface, Eqs. (\ref{eq: gluing_derivative}) and (\ref{eq: gluing_function}), to the general solution given in Eq. (\ref{eq:generic_solution}). This yields a set of two coupled nonlinear equations for $b$ and $k$ (See Appendix \ref{app:boundary}) that are invariant under the changes $V_g\to-V_g$, $\eta\to -\eta$, and $q\to -q$, provided one simultaneously replaces $k\to -k$. Therefore, once the problem has been solved for $V_g>0$, the solution for $V_g<0$ can be obtained by applying these transformations. In the following, we restrict our analysis to the case $V_g>0$. From this set of equations we obtain the following equation for $k$:
\begin{equation}
    \sn\left[\frac{k}{2\xi_c}\left(d - q X_k\right)\right]+k\dn\left[\frac{k}{2\xi_c}\left(d - q X_k\right)\right]=\frac{\sqrt{2}\xi_c}{\xi_g},
    \label{eq: k equation}
\end{equation}
where $\sn(u|m)=\sin\bm{(} \am(u|m)\bm{)}$ and $\dn(u\vert m)=\frac{d}{du} \am(u|m)$. Since both $\dn(u\vert m)$ and $\sn(u\vert m)$ are periodic functions with period $2K(m)$ and $4K(m)$ respectively, we can replace $q$ in the second equation by
\begin{equation}\label{eq:mod}
r=q\mod4.
\end{equation}
In order for the system of equations to be solvable, $q$ must satisfy the bounds:
\begin{equation}
    \frac{d}{X_k}-2<q<\frac{d}{X_k}+2.
\label{eq: q cotes}
\end{equation}
These inequalities imply that the set of values of $q$ for which the system is solvable is strongly reduced and only four values of $q$ are admissible. This is consistent with Eq. (\ref{eq:mod}). For a given $r$, the system is solvable for a single value of $\eta$. See Appendix \ref{app:boundary} for a more detailed discussion.

As descibed in Sec. \ref{sec:eqs_motion}, substituting the general solution (\ref{eq:generic_solution}), into the classical limit of $H$ in Eq. (\ref{eq:H}), we obtain the analytical expression for the dimensionless ground-state energy:
\begin{align}
    \epsilon_\text{gs} = & -\sqrt{8} \frac{\xi_c}{\xi_g}\left\{ \am \left[\frac{k}{2\xi_c}\left(d- q X_k\right)\right]+ \am \left[\frac{k}{2\xi_c} q X_k\right]\right\}\nonumber \\&+
    \frac{k^2}{2} \frac{d}{\xi_c} + 2\left[ 1 - \tanh\left(\frac{d/2-b}{\xi_c}\right)\right] \nonumber \\&+2k\left\{ \mathcal{E}\left[\frac{k}{2d}(d-qX_k)\right]+\mathcal{E} \left[\frac{k}{2\xi_c}qX_k\right]\right\}, 
    \label{eq: renormalized energy}
\end{align}
where we have defined $\epsilon_\text{gs} = \frac{2\pi \xi_c K_c}{v_c}E_\text{gs}$, and where $\mathcal{E}(u|m):=E\bm{(}\am(u|m)\bm{)}$ with $E(m)$ the complete elliptic integral of second kind. For simplicity we have omitted the second parameter $-k^{-2}$ in the argument of the elliptic functions. Interestingly, once  the integration constants $k$ and $b$ are fixed using the boundary conditions [see Eqs. (\ref{eq: k equation}) and (\ref{eq: b equation})], note that the ground-state energy becomes a universal function of the lengthscale ratios $\xi_c/\xi_g$ and $d/\xi_c$ alone.

Finally, we briefly comment on general aspects of the general solution described above. Using the condition  $\phi_c(L/2)=0$, from Eq. (\ref{eq:phi_c_minus_L_2}) we obtain the result $\phi_c(-L/2)=q \pi/\sqrt{2}$ at $x=-L/2$, which allows an explicit topological classification of our solutions. More explicitly, solutions sharing the same value of $q$ belong to the same topological sector and therefore share an identical induced charge at the gate (i.e., $Q=-eN_c=-eq$) despite small variations in the parameters $k$ and $b$ originated in, e.g.,  small differences in gate voltage $V_g$.  Moreover, it is straightforward to see that solutions belonging to the topological sector $q$ must have exactly $q$ DWs in the lattice dimerization field $\Delta(x)$. This property can be  straightforwardly seen from Eq. (\ref{eq: EOM delta}) since the function 
$\Delta(x)=-\Delta_0 \cos[\sqrt{2}\phi_c(x)]$ has exactly $q$ zeros in the segment $0\leq \phi_c(x) \leq q\pi/\sqrt{2}$. This relation between the number of bound charges and the number of DWs is a generalization of the Jackiw-Rebbi case \cite{Jackiw76_Jackiw_Rebbi_soliton, Brazovskii78_Solitons_in_Peierls_dimerized_1D_systems} (which can be associated with the case $q=1$ in our work) and has been found previously in other 1D systems, such as Bogoliubov-de Gennes and chiral Gross-Neveu systems with multiple-kink solutions \cite{Takahashe12_Self-consistent_multiple_kink_BdG_solutions}. We emphasize, however, that in contrast with the SSH model with particle-hole and inversion symmetries, which belongs to the topological class $\mathbb{Z}_2$ \cite{Asboth16_Short_course_on_TIs, Ryu10_Topological_classification, Kitaev_TI_classification}, the system considered here is crucially different due to the presence of a gate voltage which locally breaks the particle–hole symmetry, and which places the system in the topological class $\mathbb{Z}$. In the next Section we provide a formal derivation of a $\mathbb{Z}$ topological invariant and discuss these symmetry aspects in detail.

\section{Symmetry aspects and derivation of a $\mathbb{Z}$ topological invariant}\label{sec:topology}

In this Section we focus on the definition of the topological invariant that allows the classification of the interacting ground states of the system. Here we aim at providing a unified picture capable of describing our system as well as other 1D Hamiltonians falling the BDI symmetry-class, i.e., systems with time-reversal and particle-hole symmetry \cite{Altland97_Symmetry_classes, Ryu10_Topological_classification}. In seminal works on 1D systems Resta and Sorella \cite{Resta98_Position_operator_in_extended_systems, Resta99_Electron_localization} identified the (charge) Berry phase of a periodic 1D system of $N$ particles defined on a lattice as
\begin{align}
 \gamma_c&=\lim_{L\to \infty} \text{Im} \ln \langle g| e^{i(2 \pi /L )X}| g \rangle \qquad (\text{mod}\  2\pi), \label{eq:charge_berry_phase}   
\end{align}
where $|g\rangle$ is the ground state of the system, and $X=\sum_j ja(n_{j\uparrow}+ n_{j\downarrow})$ is the total position operator, with $a$ the lattice parameter and $n_{j \sigma}$ the number operator at site $j$ with spin projection $\sigma$. These works enabled a well-defined concept of localization and polarization in extended 1D systems in terms of the Berry phase in  Eq. (\ref{eq:charge_berry_phase}), i.e., $P=e \langle X\rangle/ L = e\gamma_c/(2\pi)$. Since the Berry phase $\gamma_c$ is naturally a quantity defined $\mod 2 \pi$, it became clear that $P$ could only be defined modulo $e$, consistent with the translational symmetry of the extended lattice. Additionally, in systems with inversion symmetry, the Berry phase can only take quantized values, i.e., either $\gamma_c=0$ or $\gamma_c=\pi$, corresponding to the only two inversion-symmetric points $\langle X \rangle=0$ and $\langle X \rangle =a/2 \mod a$, respectively, within a unit cell. This result allowed the definition of a $\mathbb{Z}_2$ topological invariant for the SSH model, in which the value $\gamma_c=0$ corresponds to the topologically trivial phase, and $\gamma_c=\pi$ corresponds to the nontrivial phase with localized topological edge-states in the case of open boundary conditions \cite{Asboth16_Short_course_on_TIs, Bernevig_book_TI_TSC,Shen2012_Topological_Insulators}. Finally, it was realized that expression Eq. (\ref{eq:charge_berry_phase}) reduces to the usual definition of the Zak's phase in the case of non-interacting insulating systems \cite{Zak89_Berry_phase_for_energy_bands_in_solids, Resta98_Position_operator_in_extended_systems, KingSmith93_Theory_of_polarization_in_crystalline_solids}, allowing a theoretical connection with modern topological band-theory \cite{Asboth16_Short_course_on_TIs, Bernevig_book_TI_TSC,Shen2012_Topological_Insulators}.

Interestingly, the definition of the Berry phase Eq. (\ref{eq:charge_berry_phase}) can be naturally extended to the case of strongly interacting systems \cite{Ortiz94_Many_body_formulation_of_polarization, Aligia99_Position_operator_in_extended_systems, Aligia00_Phase_diagrams_from_topological_transitions,Aligia05_Dimerized_ionic_Hubbard,Torio06_Generalized_ionic_Hubbard}. 
In particular in Ref. \cite{Aligia05_Dimerized_ionic_Hubbard}, Aligia and Batista proposed a bosonized version of Eq. (\ref{eq:charge_berry_phase}), in which the operator $U^c_L=e^{i(2\pi/L)X}$ was expressed as $U^c_L=e^{i \pi w_c}$, with $w_c$ defined as
\begin{align}
w_c &=\lim_{L\to \infty} \frac{\sqrt{8}}{\pi L} \int_{-L/2}^{L/2} \phi_c(x)\  dx,\label{eq:w_c}
\end{align}
so that substituing this expression into Eq. (\ref{eq:charge_berry_phase}) produces $\gamma_c=\pi w_c\ (\text{mod } 2\pi)$. Note that, due to the canonical commutation relations of the bosonic fields Eq. (\ref{eq:commutations}),
the operator $e^{i\sqrt{8}/L \int dx\ \phi_c(x)}$ is actually the operator generating 
the global momentum shift $\Pi_c(x)\to \Pi_c(x)-\frac{\sqrt{8}}{L}$ on the entire electronic system, precisely the effect of the operator $\exp{\left[i(2\pi/L)X\right]}$ on the ground state $|g\rangle $\cite{Resta98_Position_operator_in_extended_systems,Aligia99_Position_operator_in_extended_systems}. Since in the abovementioned references the studied systems preserve particle-hole symmetry [i.e., $\phi_c(x)\to -\phi_c(x), \ \theta_c(x)\to -\theta_c(x)$, see Ref. \cite{Montorsi17_SPT_phases_in_1D_systems_with_spin_charge_separation}] the  quantity $w_c$ is constrained to take the values 0 or 1, thus recovering the above results for $\gamma_c$. In our system, this can be easily seen in the case $V_g=0$ in the Hamiltonian (\ref{eq:H_eg}), since in that case [and assuming the spin field $\phi_s(x)$ remains always in the trivial configuration $\phi_s(x)=0$], the field $\phi_c(x)$ acquires a uniform value $\phi_c(x)=0$ or $\phi_c(x)=\pi/\sqrt{8}$, depending on the sign of the mass term $\Delta_0 - g_s/(\pi a)$. This gives rise to a $\mathbb{Z}_2$ topological invariant.

However, in the more general context systems locally violating particle-hole symmetry, but still preserving inversion symmetry (as our system in the presence of a gate), the quantity $w_c$ is promoted to a $\mathbb{Z}$ topological invariant (i.e., a winding number), allowing the topological classification of the interacting ground states. To show this property, we generically express the field $\phi_c(x)$ in its mode expansion as \cite{Haldane81_Harmonic_fluid_approach,Haldane82_Harmonic_fluid_approach_erratum, Giamarchi2004}
\begin{align}
\phi_c(x)&= \phi_0 - \frac{\pi}{\sqrt{2}L}\left(x-\frac{L}{2}\right)N_c \nonumber \\ 
 &-i\frac{\pi}{L} \sum_{q\neq 0} \sqrt{\frac{L}{\pi |q|} }\text{sign}(q) e^{-\alpha |q|/2}e^{-iq x} \left(b^\dagger_q + b_{-q}\right), \label{eq:phi_c_mode_expansion} 
\end{align}
where the boson operators $b_q, b^\dagger_q$ obey the usual commutation relations $[b_q,b_{q^\prime}^\dagger]=\delta_{q,q^\prime}$ with $q=2\pi k/L$  ($k \in \mathbb{Z}$). Substituing Eq. (\ref{eq:phi_c_mode_expansion}) into Eq. (\ref{eq:w_c}), it is straightforward to see that the term in the last line vanish upon integration due to the periodic factors $e^{-iqx}$, and we obtain 
\begin{align}
w_c&=\frac{\sqrt{8} \phi_0}{\pi} + N_c,\label{eq:top_invariant}
\end{align}
which coincides with the results in Ref. \cite{Aligia05_Dimerized_ionic_Hubbard} when $N_c=0$, precisely the particle-hole symmetric case. Interestingly, in the case $N_c=0$ we can redefine $\phi_0 \rightarrow -\phi_0$ and $b_q \to -b_q$, consistent with the particle-hole symmetry $\phi_c(x) \to -\phi_c(x)$. However, when this symmetry is broken due to a finite $V_g$, integer values $N_c \neq 0$ are allowed in the above expansion and the mapping $\phi_c(x) \rightarrow -\phi_c(x)$ is no longer possible. However, the formula for $w_c$ still yields a quantized number which can classify systems with a different number of particles. Note that our physical situation is compatible with the choice $\phi_0=0$ due to the gauge-fixing condition $\phi_c(L/2)=0$. Finally, substituing the generic symmetry condition of our solutions, Eq. (\ref{eq:symmetry_constraint}) into Eq. (\ref{eq:w_c}) we obtain $w_c=q$ with $q \in \mathbb{Z}$. Comparing this with the generic expression of $w_c$ in Eq. (\ref{eq:top_invariant}) yields the result $N_c=q$ derived in the previous Section. Therefore, we have shown that the quantity $w_c$ is a winding number invariant enabling the topological classification of our solutions.

\section{Numerical Results}\label{sec:results}

In this Section we show our main results, obtained solving Eq. (\ref{eq: k equation}) numerically for fixed $q$ (modulo 4), $\xi_c/\xi_g$ and $d/\xi_c$, and then determining the final value of $q$ using the bounds in Eq. (\ref{eq: q cotes}). 
Once these parameters are found, the ground-state energy was finally obtained using Eq. (\ref{eq: renormalized energy}). 

\begin{figure}[t]
    \centering
    \includegraphics[width=\columnwidth]{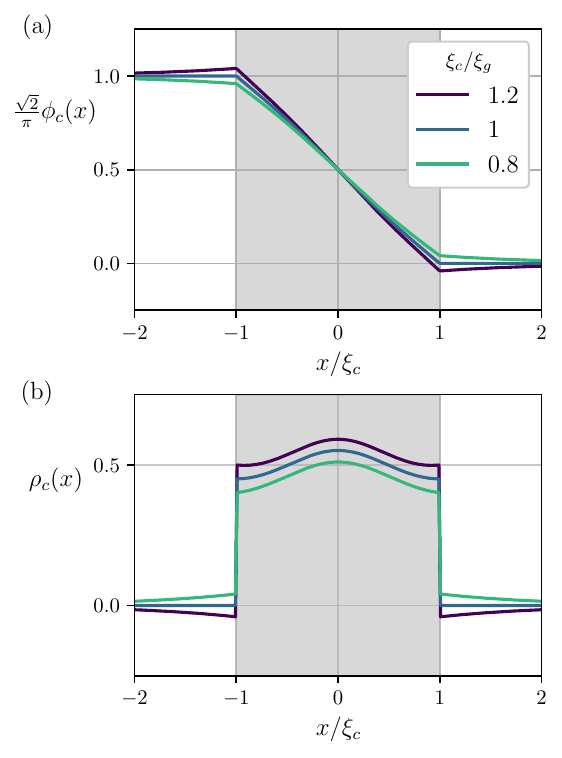}
    \caption{(Color online) Solutions in the topological sector $q=1$ (induced charge $Q=-e$). (a) Profiles of three different solutions of Eq. (\ref{eq:EOM phic final}) obtained for  fixed $d=2\xi_c$ and for different values of $\xi_c/\xi_g$. (b) The associated charge densities $\rho_c(c)$. In both plots, the gray-shaded region denotes the region $R_2$ (gated region). By smoothly varying the gate potential $V_g$ the shape of the solutions changes smoothly, but the topological properties remain unchanged. The dataset is available in Ref. \cite{dataset}}
    \label{fig: KAKtoTATtoAAA}
\end{figure}

We first illustrate the topological robustness of the solutions within a given interval of $V_g$ with a concrete example. In Fig. \ref{fig: KAKtoTATtoAAA}(a) we show the profile of $\phi_c(x)$ for fixed $d/\xi_c=2$ and for three different values of the gate voltage $V_g$, encoded in three different values $\xi_c/\xi_g$ via Eq. (\ref{eq:xi_g}). In Fig.  \ref{fig: KAKtoTATtoAAA}(b) we show the charge density, obtained by deriving $\phi_c(x)$ with respect to $x$ [see Eq. (\ref{eq:charge_density})].
Despite the small variations in shape,
the three curves belong to the same topological sector $q=1$ (total induced charge  $Q=-e$). This can be more intuitively seen in Fig. \ref{fig: KAKtoTATtoAAA}(b), where although the charge profile changes in response to the small variations in $V_g$, the total area below each curve remains constant, meaning that the total charge remains constant and perfectly quantized according to the formula $Q=-eq$, despite these changes in $V_g$. In addition, note that the charge is essentially localized at the region of the gate within a lengthscale $\xi_c$. These two features confirm that robust charge quantization at the gate can be achieved in this device.

\begin{figure*}[t!]
    \centering    \includegraphics[width=2\columnwidth]{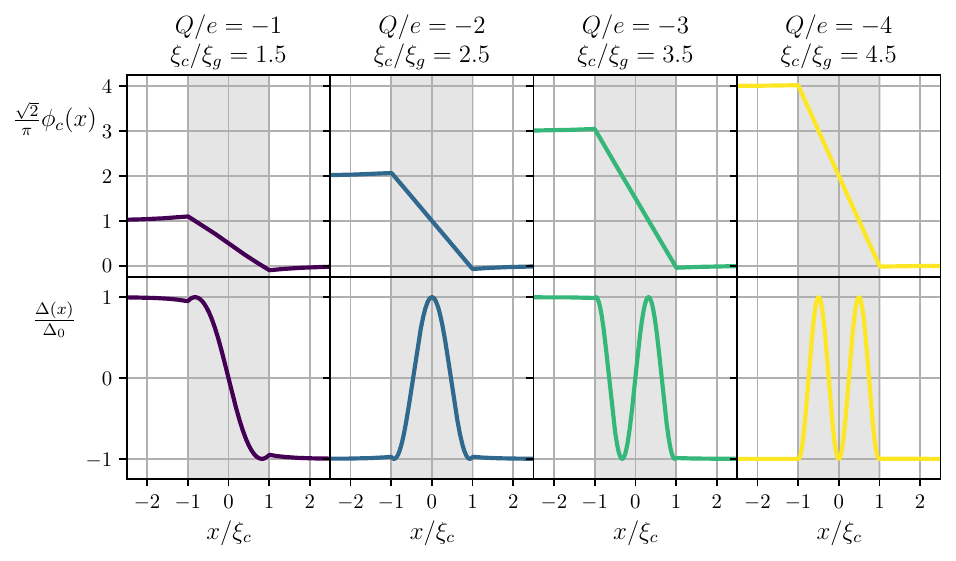}
    \caption{(Color online) Ground-state configuration of the field $\phi_c(x)$ and the associated (staggered) lattice dimerization field $\Delta(x)$, obtained for different values of the gate potential $V_g$. As $V_g$ is increased, a sequence of topological phase transitions are produced, in which the value $\phi_c(-L/2)$ jumps in units of $\sqrt{2}/\pi$ while simultaneously a new DW appears in the gated region (gray-shaded areas). The dataset is available in Ref. \cite{dataset}}
    \label{fig: phi and delta}
\end{figure*}

By increasing $V_g$ (and hence the parameter $\xi_c/\xi_g$) further, a sequence of topological transitions is induced at specific critical values, separating distinct topological sectors characterized by integer values $q = 0, 1, 2, 3,\dots$. As mentioned before, configurations belonging to different topological sectors exhibit distinctive qualitative features. In addition to the quantized electronic charge, these features can also be clearly visualized in the dimerization field $\Delta(x)$ which, under the assumption of no excitations in the spin sector (i.e., $S = 0$ in the ground state), is fully determined by $\phi_c(x)$ through Eq.~(\ref{eq: EOM delta}). In Fig.~\ref{fig: phi and delta}, we present solutions from different topological sectors, selected by choosing representative values of $\xi_c/\xi_g$, and for fixed $d/\xi_c=2$. The top row shows the field profile $\phi_c(x)$, while the bottom row displays the corresponding $\Delta(x)$.
As $V_g$ increases and since we fixed $\phi_c(L/2)=0$, the topological phase transitions manifest as discrete jumps of magnitude $\sqrt{2}/\pi$ in $\phi_c(-L/2)$, accompanied by the emergence of an additional DW within the gated region. As already mentioned, the number of DWs (or equivalently, the number of zeros of $\Delta(x)$) exactly corresponds to $q$, reflecting the topological charge of the multikink configurations. As mentioned in Sec. \ref{sec:theoretical_framework}, note that solutions with even $q$ connect the same values of $\Delta(x)$ at $x=\pm L/2$ ($-\Delta_0$ to $-\Delta_0$ for $q=2$ and $4$ in Fig. \ref{fig: phi and delta}) consistent with the periodic boundary conditions. On the other hand, solutions with odd $q$ connect  $\Delta_0$ and $-\Delta_0$ (cases $q=1$ and 3 in Fig. \ref{fig: phi and delta}) and therefore, correspond to antiperiodic boundary conditions in the field $\Delta(x)$. 

\begin{figure}
    \centering
    \includegraphics[width=\columnwidth]{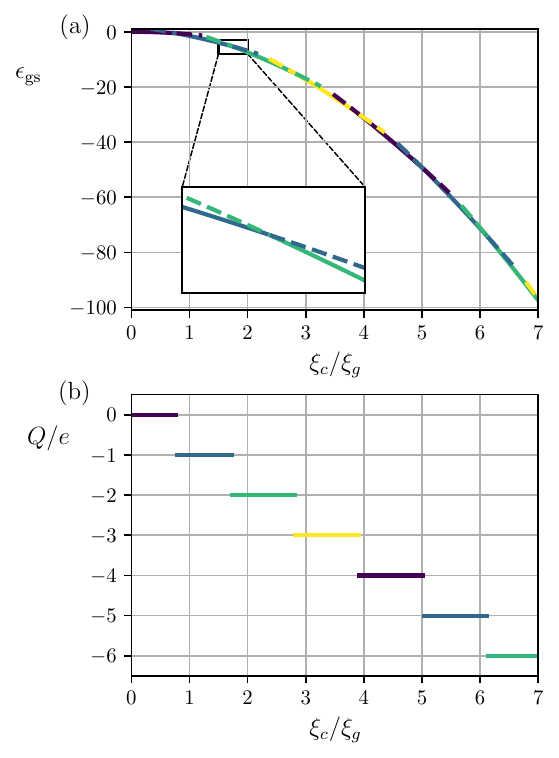}
    \caption{(Color online) (a) Dimensionless ground-state energy $\epsilon_\text{gs}$ and (b) total induced charge $Q$ as functions of the ratio $\xi_c/\xi_g \sim V_g$ for fixed $d=2\xi_c$. In panel (a), the solid lines correspond to the ground-state energy in a given topological sector, whereas the dashed lines correspond to the first excited state. Topological quantum phase      transitions occur at critical voltages where energy levels cross. Whenever this transition occurs,the total charge $Q$ changes jumps by $-e$. The dataset is available in Ref. \cite{dataset}}
    \label{fig: eGS}
\end{figure}

To better understand the origin and nature of the topological transitions, we now focus on the dimensionless ground-state energy (\ref{eq: renormalized energy}).
In Fig. \ref{fig: eGS}(a) we plot  $\epsilon_\text{gs}$ as a function of $\xi_c/\xi_g$ for fixed $d/\xi_c=2$, and in Fig. \ref{fig: eGS}(b) we show the corresponding electronic charge induced at the gate. 
The region of $V_g$ where a particular solution corresponds to the ground state defines a given topological sector $q$. In this region, the corresponding ground-state energy $\epsilon_\text{gs}$ is plotted with solid lines. At the critical points, level-crossings occur where an excited state from a different topological sector, indicated here by a different color, becomes the new ground state. In dashed lines we have shown the extrapolation of a given ground-state solution into a region where it has become an excited state (i.e., a topological excitation). In the inset of Figure \ref{fig: eGS}(a) we show the level crossing between the topological sectors $q=1$ and $q=2$. Figure \ref{fig: eGS}(a) then allows to interpret the topological transitions as first-order ground-state level crossings driven by $V_g$. In addition, in Fig. \ref{fig: eGS}(b) we note an interesting property: the total topological charge $Q$ induced at the gate is essentially  the derivative of $\epsilon_\text{gs}(V_g)$ with respect to $V_g$ (up to small corrections due to charge-density contributions lying outside the integration limits within a distance $\approx \xi_c$). 
Then, the topological transitions and the associated jumps in $Q$ at the critical voltages share the same physical origin: the level-crossings of the ground-state energy branches and the inherent discontinuity of their derivative. This provides a  unifying intuitive picture of the underlying physics for the topological quantum phase transitions found in this work, in accordance to the findings of Ref. \cite{Arancibia25_Towards_electrical_DW_control} obtained for a non-interacting tPA molecule.  
\begin{figure}
    \centering
    \includegraphics[width=0.9\linewidth]{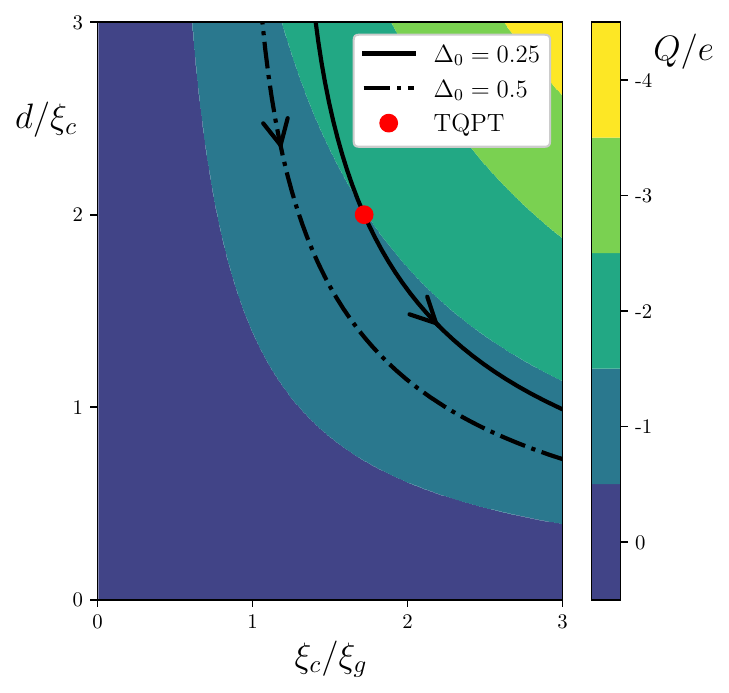}
    \caption{(Color online) Topological phase diagram in the $(d/\xi_c,\xi_c/\xi_g)$ plane. Each colored region corresponds to a different topological sector. The black line corresponds to a parametric curve generated as the parameters $d/\xi_c$ and $\xi_c/\xi_g$ get renormalized by the interaction parameter $g$ [see Eqs. (\ref{eq:xic_renormalized})--(\ref{eq:xic_xig_renormalized})], where the arrow indicates de direction in which $g$ increase. The dataset is available in Ref. \cite{dataset}}
    \label{fig: phasediagram}
\end{figure}

Finally, in Fig.  \ref{fig: phasediagram} we show the topological phase diagram of the system, expressed as a universal  function of the dimensionless quantities $d/\xi_c$, representing the width of the potential, and $\xi_c/\xi_g \propto V_g$, which quantifies its depth. The colored regions correspond to distinct topological sectors characterized by the quantized charge $Q$. As either $d$ or $V_g$ are  increased, the system undergoes sequential quantum phase transitions where the excess charge $Q$ is decreased by $-e$ at each step of a staircase-type diagram. 
In this plot, the Fig. \ref{fig: eGS}(b) presents previously actually corresponds to a horizontal cut along the axis $\xi_c/\xi_g$ made at the value of the vertical axis $d/\xi_c=2$. Similarly to the case of the infinite square-well potential seen in introductory quantum mechanics courses (e.g., Ref. \cite{Sakurai1993Modern}) and assuming the system is weakly coupled to a particle reservoir, in the non-interacting case one can intuitively see that whenever the width of the well $d$ or its depth $V_g$ increase, the total charge contained in the well must increase. However, we stress that the topological phase diagram in Fig.  \ref{fig: phasediagram} is a highly nontrivial result that goes beyond the single-particle framework and describes the topological ground-state of the fully interacting system. To illustrate the effect of repulsive interaction, we consider the particular case where all interaction parameters appearing in Eqs. (\ref{Hfs})--(\ref{Hum}) are identical, i.e., $g_{\sigma \sigma^\prime, \alpha \alpha^\prime}=g_s=g_c\equiv g$, with $g>0$. This is of course not a general situation, since the  specific values of the coupling parameters are highly dependent on the details of the original microscopic model. Moreover, we recall that Eqs. (\ref{Hfs})--(\ref{Hum}) describe purely contact interactions, corresponding to the case of infinitely screened Coulomb interactions having zero range (see Sec. \ref{sec:summary} for a further discussion). Nevertheless, our goal here is to gather useful physical intuition and not to provide a quantitative description of a real experiment. For this specific case we have the parameter relation  $v_c/K_c=v_F(1+2g/\pi v_F)$ and
\begin{align}
    \xi_c^{-2}&=\frac{4}{v_F a}\frac{\Delta_0-g/(\pi a)}{1+2g/(\pi v_F)}.\label{eq:xic_renormalized}
\end{align}
From this expression we note that as $g$ increases, so does the localization length $\xi_c$. Then, by fixing $d$ the renormalized width $d/\xi_c$ decreases as interactions become more repulsive, giving less effective space for the system to accommodate charge. On the other hand,
\begin{align}
    \xi_g = \frac{v_F}{\sqrt{2}}\left(1+\frac{2g}{\pi v_F}\right)V_g^{-1},\label{eq:xig_renormalized}
\end{align}
meaning that, for fixed $V_g$, the length-scale $\xi_g$ also increases with $g$. In other words, the  interaction $g$ renormalizes the potentia depth $V_g \to V_g/[1+2g/(\pi v_F)]$. This can be interpreted as a less effective gate potential due to the effect of electron-electron repulsion. The ratio $\xi_c/\xi_g$ is then given by
\begin{align}\label{eq:xic_xig_renormalized}
    \frac{\xi_c}{\xi_g}=\sqrt{\frac{a}{2v_F\left(\Delta_0-\frac{g}{\pi a}\right)\left(1+\frac{2g}{\pi v_F}\right)}}V_g,
\end{align}
and encodes a nontrivial competition  between electron-electron, electron-lattice, and electron-gate interactions as a function of $g$. Imposing the physical condition $\Delta_0 \ll a/v_F \sim E_F$ (i.e., a Peierls gap much smaller than the Fermi energy) we note that the ratio $\xi_c/\xi_g$ is a monotonically increasing function of $g$. Since we have assumed the case of weak interactions, we must restrict the value of the interaction parameter to $g< \Delta_0 \pi a$ for the consistency of our results. Note that as we approach the limit $g\to \Delta_0 \pi a$, we obtain $d/\xi_c\to0$ and $\xi_c/\xi_g\to\infty$. In this limit the curves delimiting the different topological sectors collapse to horizontal axis and the charge $Q$ becomes ill-defined. 

To illustrate the behavior of the system as a function of $g$, in Fig.~\ref{fig: phasediagram} we show two representative trajectories in parameter space with initial conditions $V_g = E_F$ and $d = 3a$, and two different values of the Peierls gap: $\Delta_0 = 0.25 E_F$ (solid black line) and $\Delta_0 = 0.5 E_F$ (dashed black line). As $g$ increases from zero, the ground state evolves following the flow of the renormalized ratios $d/\xi_c(g)$ and $\xi_c/\xi_g(g)$. In particular, the solid curve exhibits an interaction-induced topological quantum phase transition (TQPT) between the topological sectors $q=2 \to 1$, occurring at $\xi_c/\xi_g \approx 1.7$ and $d/\xi_c \approx 2.0$, for a critical interaction strength $g = 0.37 a$. This TQPT, in which the system becomes discharged, is physically expected in the presence of repulsive interactions. However, we emphasize that this behavior is not universal: with suitable fine-tuning, the system may instead undergo the opposite transition. We attribute this counterintuitive result to the assumption of purely contact-type interactions in our model, which enhances the competition between the underlying physical mechanisms and renders the phase boundaries particularly sensitive to parameter variations. We therefore expect that the inclusion of longer-range interactions would lead to significant modifications of the topological phase diagram. Finally, for a different set of parameters (dashed curve), the system can be tuned into a stable charge plateau, with no topological transition within the explored parameter regime.

\section{Summary and Conclusions}\label{sec:summary}

We have theoretically investigated the ground-state properties of a linear trans-polyacetylene molecule capacitively coupled to an external gate voltage 
$V_g$. Despite the simplicity of the model, we have shown that this system exhibits novel topological properties of interest both for fundamental research and for potential applications in organic nanoelectronics. The system is modeled using the Takayama-Lin-Liu-Maki (TLM) model, the continuum limit of the Su-Schrieffer-Heeger (SSH) Hamiltonian, and analyzed within the framework of Abelian bosonization. Both electronic and lattice degrees of freedom are explicitly considered, which is essential for determining the equilibrium ground-state configuration and for capturing effects arising from the interplay of topology, electron-electron interactions, and the applied gate voltage.

Within the bosonization formalism, the interacting ground state is described by multikink solutions of a modified sine-Gordon equation for the charge-density field [see Eqs. (\ref{eq:EOM phic final}) and (\ref{eq:EOM_phic_final_dimensionless})]. These equations and their solutions constitute central results of this work. The multikink solutions depend on the gate voltage $V_g$, the gate width $d$, and the Luttinger interaction parameter $K_c$, and are classified into topological sectors labeled by an integer $q$, which acts as a topological invariant characterizing both the induced electronic charge in the gated region (i.e., $Q=-eq$) and the number of induced DWs. In agreement with a previous study that focused on the noninteracting limit \cite{Arancibia25_Towards_electrical_DW_control}, increasing $V_g$ leads to a sequence of topological phase transitions, signaled by abrupt changes in $q$. However, the present work goes beyond the non-interacting limit and provides a unified framework that also captures interacting systems. The bosonization formalism naturally extends previous results by incorporating electron-electron interactions and, in particular, shows that since these interactions do not break the protecting symmetries of the model, they do not qualitatively affect the topological phases. An important feature emerging from the interacting picture, however, is the possibility of interaction-induced topological phase transitions, which is absent in Ref. \cite{Arancibia25_Towards_electrical_DW_control}. In addition, within the bosonization formalism we obtain the analytical form of a $\mathbb{Z}$ topological invariant (i.e., a winding number), which is crucial for the classification of the interacting ground states.  Another key feature of the present work is that, in contrast with standard theoretical studies of the SSH model (see e.g., Ref. \cite{Asboth16_Short_course_on_TIs}), electronic and lattice degrees of freedom must be considered self-consistently. Minimization of the classical ground-state energy subject to the self-consistency condition Eq.~(\ref{eq: EOM delta}), yields qualitatively new multikink solutions in the presence of an applied gate voltage.

Moreover, we have also investigated the effect of a weak repulsive Coulomb interaction (i.e., obeying the weak-coupling condition $\{g_{\sigma \sigma^\prime, \alpha \alpha^\prime}, g_s, g_c\} \ll a \Delta_0$) on the ground-state configurations and on the resulting topological phase diagram. Within the bosonization framework correlations can be incorporated in a straightforward yet nontrivial manner. Specifically, interaction effects are fully encoded in the analytical expressions of $\xi_c$ and $\xi_g$, and result in interaction-induced renormalizations of these  characteristic lengthscales. This allows a universal characterization of the interacting system in terms of the ratios $d/\xi_c$ and $\xi_c/\xi_g$. For fixed gate width $d$ and gate voltage $V_g$, repulsive interactions renormalize these ratios and can eventually drive interaction-induced topological phase transitions (see Fig. \ref{fig: phasediagram}). In addition, our analytical results allow us to study the connection between the interacting and noninteracting cases in the limit of a vanishing interaction parameter $g\rightarrow 0$.

Despite the simplifying assumptions and idealizations adopted in our model, such as a square-shaped potential generated by the gate and the neglect of disorder, we expect our central result, namely the possibility of inducing topological phase transitions in a tPA molecule via an external gate voltage, to be qualitatively robust and potentially relevant for the emerging field of topological quantum devices. From a practical standpoint, our results indicate that the topological charges induced in gated tPA molecules would be robust and stable against small variations of the gate potential $V_g$. Extracting parameters from the discrete SSH model so that it reproduces the experimental value of the bulk Peierls gap $\Delta_0 \approx 0.7$ eV (see e.g.,  Ref. \cite{ono1990motion}), we obtain critical values $V_g^{(q)}$ of the order of 0.1 V, which is within experimentally reasonable values. Of course, this would also depend on the details of the capacitive coupling of the gate to the tPA molecule. But besides these experimental details, the robustness of the charge quantization could be exploited in concrete nanoelectronic applications, e.g., the possibility to fabricate topologically protected organic quantum dots (QDs) harboring perfectly quantized charges. Note that in conventional Coulomb-blockaded semiconductor QDs the charge-quantization mechanism is a purely local phenomenon, governed by the local Coulomb repulsion (e.g., the parameter $U$ in the Anderson model) and by the absence of electronic levels near the Fermi energy within the level-broadening scale $\gamma$. By contrast, here charge quantization is an intrinsically topological phenomenon that involves the entire system. A simple way to see this distinction is that the emergence of a gate-induced DW requires all bonds to the left of the gate to switch from long to short, or vice versa. Thus, the presence of a DW (and the ability to accommodate an additional charge at the gate) is not a local but a collective topological phenomenon.

Another approximation we have introduced concerns the range of the electron-electron interaction. Specifically, our model applies in the regime of strongly screened Coulomb interaction between electrons (for instance, due to a nearby metallic surface) so that the potential effectively has a finite range and can be replaced by a Dirac-delta potential in the long-wavelength limit. When screening is less effective and electrostatic interactions are therefore stronger, we anticipate modifications in the detailed structure of the phase diagram. Nevertheless, we still expect the presence of the topological phase transitions.

While it would be experimentally hard to infer the accumulated charge $Q$ near the gate directly from electrical measurements,  the value of the topological invariant $q$ (and therefore, the quantized charge via the relation $Q=-eq$) could be indirectly inferred using high-resolution atomic-force microscopy (AFM) or STM. By analyzing the positions of the ions in the tPA molecule in the vicinity of the gate, the dimerization pattern $\Delta(x)$ and its number of nodes could be in principle extracted. The possibility of directly resolving bond-alternation lengths in a single tPA molecule using AFM has been experimentally  demonstrated recently in Ref. \cite{Wang19_Solitons_in_individual_PA_molecules}. Given the rapid progress in on-surface synthesis techniques, the maturity of STM or AFM methods for probing nanostructured systems, and the availability of multiple approaches for fabricating electronic nanocircuits, we believe that experimental realization of the proposed device lies within current technological capabilities.

Finally, we note that, due to the explicit symmetry of the bosonization formalism under the interchange of the charge and spin sectors [i.e., the interchange of the fields $\phi_c(x) \leftrightarrow \phi_s(x)$ and, simultaneously, the couplings $g_c \leftrightarrow g_s$], a magnetic analog of the presented results could be obtained by replacing the voltage gate with a localized Zeeman exchange field of width $d$. Independently of the associated experimental challenges, such a model could be physically realized by placing a narrow, finger-shaped segment of a ferromagnetic insulating material (such as EuS) on top of the tPA molecule \cite{Tedrow86_Spin_polarized_transport_EuO_Al, Diesch18_Equal_spin_triplet_superconductivity_at_EuS_Al_interfaces}. Then, if the proximity-induced exchange field $h$ is sufficiently large, we predict the occurrence of topological transitions with induced local spins and the concomitant generation of magnetic DWs, enabling novel and interesting effects with potential interest in the field of magnetic organic nanomaterials.

\acknowledgments

The authors are grateful to A. A. Aligia for helpful discussions. This work was partially supported by UNLP under Grant PID X497. 

\section*{Data Availability}

The data that supports the findings of this article are openly available \cite{dataset}.

\appendix

\section{Interacting electrons in the Abelian bosonization formalism}\label{app:fermions}

The Hamiltonian of one-dimensional (1D) electrons at low energy is typically written as: $H_\text{e}=H_0+H_e^\text{fs}+H_e^\text{bs}+H_e^\text{um}$, where
\begin{align}
H_0 &= -i v_\mathrm{F} \sum_\sigma \int_{-L/2}^{L/2} dx \,\left( \psi_{\sigma  R }^\dagger \partial_x \psi_{\sigma  R } - \psi_{\sigma L }^\dagger \partial_x \psi_{\sigma L }\right), \label{H0} \\
  H_e^\text{fs} &= \sum_{\sigma\sigma'}\sum_{\alpha\alpha'} g_{\sigma\sigma',\alpha\alpha'} \int_{-L/2}^{L/2} dx \, \rho_{\sigma \alpha }\left(x\right)\rho_{\sigma' \alpha' }\left(x\right), \label{Hfs}\\
H_e^\text{bs} &= g_s\sum_\sigma \int_{-L/2}^{L/2} dx\ \psi_{L\sigma}^\dagger \psi_{R \sigma} \psi_{R\bar{\sigma}}^\dagger\psi_{L\bar{\sigma}},\label{Hbs}   \\
  H_e^\text{um} &= \frac{g_c}{2} \sum_{\sigma}  \int_{-L/2}^{L/2} dx\, \left(\psi_{\sigma L}^{\dagger}\psi_{\bar{\sigma} L}^{\dagger}\psi_{\bar{\sigma}R}\psi_{\sigma R}+\hc\right).
  \label{Hum}
\end{align}
with $\psi_{\sigma\alpha}$ being the annihilation operators of chiral fermion fields with chirality $\alpha=-\bar{\alpha}\in\{R,L\}\rightarrow\{\pm 1\}$ and spin $\sigma=-\bar\sigma\in\{\uparrow,\downarrow\}\to\{\pm1\}$. The term $H_0$ in Eq.~(\ref{H0}) represents the electronic kinetic energy obtained from the usual linearization of the band structure around the Fermi points $\pm k_F$. The Fermi velocity is obtained from the relation $v_F=\hbar^{-1} \left.\partial \varepsilon_k /\partial k \right|_{E_F}$, where $\varepsilon_k$ is the conduction-electron dispersion relation. Note that here we have adopted the convention $\hbar=1$

The term $H_e^\text{fs}$ in Eq. (\ref{Hfs}) represents forward-scattering interactions, where we have defined the fermionic density of each fermion species as $\rho_{\sigma\alpha}(x) = \ \psi^\dagger_{\sigma\alpha}(x)\psi_{\sigma\alpha}(x)$. In the presence of inversion and SU(2) symmetry, as will be the case in this work, the Coulomb interaction parameters obey many symmetry constraints and reduce to four parameters defined (using the notation in Ref. \cite{Giamarchi2004}) as 
\begin{equation}
\begin{aligned}
g_{4\parallel}&= g_{\sigma \sigma,\alpha \alpha},\\ g_{4\perp}&= g_{\sigma \bar{\sigma},\alpha \alpha},\\ 
g_{2\parallel}&= g_{\sigma \sigma,\alpha \bar{\alpha}},\\ 
g_{2\perp}&= g_{\sigma \bar{\sigma},\alpha \bar{\alpha}}.
\end{aligned}\label{eq:geology}
\end{equation}

The term $H_e^\text{bs}$ in Eq. (\ref{Hbs}) represents backward-scattering interactions where electrons exchange dispersion branches while keeping the total momentum exchange equal to zero \cite{Giamarchi2004, Gogolin1998}. For compactness in our notation, we have dropped the position argument in the fermionic fields in Eqs. \ref{H0}--\ref{Hum}, i.e., $\psi^\dagger_{\sigma \alpha}(x)\equiv \psi^\dagger_{\sigma \alpha}$.

The case of an undoped (i.e., neutral) tPA chain corresponds to half-filling of the electronic band where the Fermi momentum is $k_F=\pi/2a$. Then, the so-called ``umklapp" scattering terms, represented by the Hamiltonian $H_e^\text{um}$ in Eq. (\ref{Hum}) and describing processes where two right movers scatter into two left movers, or vice versa, transferring a net momentum $4k_F=2\pi/a$ to the lattice, must be kept in the low-energy description \cite{Giamarchi2004, Gogolin1998}.

Within the  Abelian bosonization technique \cite{Giamarchi2004,Gogolin1998}, the chiral fermions are represented as 
\begin{align}
    \psi_{\sigma\alpha}(x)&=\frac{\eta_{\sigma\alpha}}{\sqrt{2\pi a}}e^{-i\alpha\phi_{\sigma\alpha}(x)},\label{eq:bosonization_formula}
\end{align}
where $\phi_{\sigma\alpha}(x)$ are chiral bosonic fields obeying the Kac-Moody commutation relations $[\phi_{\sigma\alpha}(x),\phi_{\sigma'\alpha'}(x')] = i\pi \alpha\delta_{\alpha\alpha'}\delta_{\sigma\sigma'}\sign(x-x')$, $a \sim k_F^{-1}$ is the short-distance cutoff of the continuum theory (assumed to correspond to the lattice parameter in the discrete SSH model). The operators $\eta_{\sigma\alpha}$ are the standard Majorana fermions that ensure the proper anticommutation relations of the fermionic chiral fields. In addition, the chiral densities are written in bosonized form as $\rho_{\sigma\alpha}(x)=-\partial_x\phi_{\sigma\alpha}(x)/2\pi$.

Following standard procedures, we now introduce spin $s$ and charge $c$ dual fields through the transformation  \cite{Giamarchi2004,Gogolin1998}
\begin{align}
\sqrt{2} \phi_{\sigma\alpha}(x) &= \phi_c(x) + \sigma \phi_s(x) -\alpha \theta_c(x) - \sigma \alpha \theta_s(x),\label{eq:field_transformation}
\end{align}
where the convention  $\sigma=+1(-1)$ for $\sigma=\uparrow (\downarrow)$ and $\alpha=+1(-1)$ for $\alpha=R (L)$ on the right-hand side is implied. Introducing $\Pi_\nu=\partial_x\theta_\nu/\pi$ one shows that they correspond to the momenta conjugate to $\phi_\nu$ and thus obey canonical commutation relations
\begin{equation}
    \left[ \phi_\nu(x), \Pi_\mu(y) \right]= i \delta(x-y)\delta_{\mu\nu},
    \label{eq:commutations}
\end{equation}
with $\{\nu,\mu\}=\{c,s\}$.

 In the bosonic language, the charge and spin densities become
\begin{align}
    \rho_c(x)&=\sum_{\sigma,\alpha}\rho_{\sigma\alpha}(x)=-\frac{\sqrt{2}}{\pi}\partial_x\phi_c(x),\label{eq:charge_density}\\
    \rho_s(x)&=\sum_{\sigma,\alpha}\sigma\rho_{\sigma\alpha}(x)=-\frac{\sqrt{2}}{\pi}\partial_x\phi_s(x).
\end{align}
 A standard manipulation after bosonization renders the kinetic and forward-scattering part of the electronic Hamiltonian into a quadratic Hamiltonian, the so-called Luttinger-Liquid model, i.e.,  $H_{LL}= H_0+H_e^\text{fs}$ \cite{Giamarchi2004, Gogolin1998}. This term naturally splits as $H_{LL}=H_{0,s}+H_{0,c}$, explicitly separating charge and spin modes,
\begin{equation}
   H_{0,\nu}=\frac{v_\nu}{2\pi}\int_{-L/2}^{L/2} dx\,\left[\frac{1}{K_\nu}(\partial_x\phi_\nu)^2+K_\nu\pi^2\Pi_\nu^2\right],
   \label{eq:Hnu}
\end{equation}
with  $\nu=\{c,s\}$. In this expression, $v_\nu$ physically represents the velocity of charge or  spin collective modes (i.e., plasmons or spinons, respectively) and $K_\nu$ are the dimensionless Luttinger parameters (or stiffness parameters) that control the decay of the correlation functions.
In terms of the forward-scattering couplings defined in Eq. (\ref{eq:geology}), $K_\nu$ are given by
\begin{align}
    v_\nu &= v_F\sqrt{(1+y_{4\nu}/2)^2-(y_\nu/2)^2},\\
    K_\nu &=\sqrt{\frac{1+y_{4\nu}/2+y_\nu/2}{1+y_{4\nu}/2-y_\nu/2}},
\end{align}
where, using the definitions in Eq. (\ref{eq:geology}), we have defined $y_{4\nu}=(g_{4||}\pm g_{4\perp})/(\pi v_F)$ and $y_\nu =(-g_{2||}\mp g_{2\perp})/(\pi v_F)$, with the upper (lower) sign for $\nu=c$ ($\nu=s$). 

Finally, the backscattering and umklapp terms are given by
\begin{align}
        H_e^\text{bs} &= \frac{2 g_s}{(2\pi a)^2}\int_{-L/2}^{L/2} dx\,\cos \sqrt{8}\phi_s(x),\\
H_e^\text{um}&=\frac{2g_c}{(2\pi a)^2} \int_{-L/2}^{L/2} dx\,\cos \sqrt{8}\phi_c(x),
\end{align}
respectively, so that backward and unmklapp scattering do not spoil spin-charge separation. The total Hamiltonian $H_e$ compactly writes as in Eq. (\ref{eq:H_e})

\section{Solution of Eq. (\ref{eq:EOM phic final})}\label{app:Jacobi}

To solve the equation of motion for $\phi_c$ [Eq. (\ref{eq:EOM phic final})], we divide the space in three regions: $R_1=(-\infty,-d/2)$, $R_1=(-d/2,d/2)$ and $R_3=(d/2,+\infty)$. In each region  we have to solve the sine-Gordon equation:
\begin{equation}
    \phi_i''(x) -\frac{\sin\sqrt{8}\phi_i(x)}{\sqrt{8}\xi^2_c}=0,
    \label{ec: sinegordon}
\end{equation}
where $i=1,2,3$ and $\phi_i$ is the solution in region $R_i$. We begin to solve in $R_1$ and $R_3$. For that purpose, we recall the boundary conditions at $x=\pm\infty$:
\begin{align}
\phi_c'(x=\pm\infty)&= 0,\label{eq:phi_prime}\\ 
\cos[\sqrt{8}\phi_c(x=\pm\infty)]&=1.\label{eq:cos_phi}
\end{align}
Note that if we find a function $\phi_i(x)$ that satisfies both Eq. (\ref{ec: sinegordon}) and Eqs. (\ref{eq:phi_prime}) and (\ref{eq:cos_phi}), then $\tilde{\phi}_i(x)=\phi_i(x)+\frac{n_i\pi}{\sqrt{2}}$ also does. We start by noting that 
\begin{equation}
    \phi_i''=\phi_i'\frac{d\phi'}{d\phi}.
\end{equation}
Taking this into account, we integrate Eq. (\ref{ec: sinegordon}) from $-\infty$ ($+\infty$) to $x$ in $R_1$ ($R_3$), and then integrate once more to obtain the well-known solitonic solutions of the sine-Gordon equation:
\begin{equation}
    \phi_c^i(x)= \sqrt{2} \arctan \left\{\exp \left[\frac{\eta_i(x-b_i)}{\xi_c}\right]\right\}+\frac{n_i\pi}{\sqrt{2}},
\end{equation}
where $b_i$ are constants of integration that must be compatible with the boundary conditions at the edge of the gate, and $\eta_i=-1,0,1$ characterize the antikink, trivial and kink solution respectively.

Using the identity
\begin{equation}
    \arctan(y)+\arctan(1/y)=\pi/2,
\end{equation}
and noting that $\phi_c$ is defined modulo $\pi/\sqrt{2}$, we can conveniently write
\begin{align}
    \phi_1(x) &= \eta_1 \sqrt{2}\arctan \left\{\exp \left[\frac{(x-b_1)}{\xi_c}\right]\right\} +\frac{n_1\pi}{\sqrt{2}},\\
    \phi_3(x)& = - \eta_3\sqrt{2} \arctan \left\{\exp \left[\frac{-(x-b_3)}{\xi_c}\right]\right\}. 
\end{align}

We now turn to the solution in $R_2$. We start by integrating Eq. (\ref{ec: sinegordon}) from $-d/2$ to $x$, obtaining
\begin{equation}
\begin{aligned}
    \frac{1}{2}\phi'^2_2(x)-\frac{1}{2}\phi'^2_2(-d/2) = \frac{1}{8\xi_c^2}\left[\cos \sqrt{8}\phi_2 (-d/2)\right.&\\
    -\left.\cos \sqrt{8}\phi_2(x)\right].
\end{aligned}
\end{equation}

By defining $k^2=2\xi_c^2\phi_2'^2(-d/2)-\sin^2 \sqrt{2}\phi_2(-d/2)$, and using standard trigonometric identities we obtain
\begin{align}
    \phi_2' = \pm \frac{1}{\sqrt{2}\xi_c}\sqrt{\sin^2 \sqrt{2}\phi_2+k^2}.
\end{align}
Integrating once more yields
\begin{align}
    \phi_{2}(x)= \frac{\eta_2}{\sqrt{2}}\am \left[\frac{k(x-b_2)}{\xi_c}\Bigg\vert-k^{-2}\right],
\end{align}
where $b_2$ is an additional integration constant and $\am (x|m)$ denotes the Jacobi amplitude, defined as the inverse of the incomplete integral of the first kind:
\begin{align}
    F(x|m)= \int_0^x \frac{dy}{\sqrt{1-k^2\sin^2y}}.
\end{align}
Here we absorbed the constant $n_2/\sqrt{2}\pi$ in $b_2$ by using the property
\begin{align}
    \am[u+2K(m)|m]=\am[u|m]+\pi,
\end{align}
where $K(m)=F(\pi/2,m)$ is the complete elliptic integral of the first kind. Note that the case $\eta_2=0$ is not considered beacuse it cannot satisfy boundary conditions at the edge of the gate. The symmetry constraints discussed in Sec. \ref{sec:eqs_motion} imply $b_3=-b_1:=b$, $\eta_1=\eta_3:=\eta$ and $b_2=\eta_2 q K(-k^2)$. The parameters $k$ and $b$ are then determined by imposing the boundary conditions at $x=\pm d/2$, which encode the effects of the gate through Eq. (\ref{eq:EOM phic final}).

\section{Integration constants from boundary conditions}\label{app:boundary}

To find the integrations constants $b$ and $k$, we impose the boundary conditions at the interface, Eqs. (\ref{eq: gluing_derivative}) and (\ref{eq: gluing_function}), to the general solution given in Eq. (\ref{eq:generic_solution}):
\begin{align}
    &2\eta \arctan\left \{\exp\left [-\frac{(d/2-b)}{\xi_c}\right]\right\} - \am \left[ \frac{k}{2\xi_c}\left(d- q X_k\right)\right] =0,\label{eq: continuity}\\
    &\eta \sech \left[ \frac{(d/2-b)}{\xi_c} \right ] + k \dn\left[\frac{k}{2\xi_c}\left(d- q X_k\right)\right] =\frac{\xi_c}{\xi_g}. \label{eq: derivability}
\end{align}
Solving Eq. (\ref{eq: continuity}) for $b$ in terms of $k$ we obtain 
\begin{equation}
    b = \xi_c\log\left\{\tan\left[\frac{\eta }{2} \am \left[\frac{k}{2\xi_c}\left(d - q X_k\right)\right]\right]\right\}+ \frac{d}{2},  \label{eq: b equation}
\end{equation}
and substituing Eq. (\ref{eq: b equation})  into Eq. (\ref{eq: derivability}), we obtain the following equation for $k$: 
\begin{equation}
    \sn\left[\frac{k}{2\xi_c}\left(d - q X_k\right)\right]+k\dn\left[\frac{k}{2\xi_c}\left(d - q X_k\right)\right]=\sqrt{2}\frac{\xi_c}{\xi_g}.
    \label{eq: k equation AP}
\end{equation}
It is important to note that not all values of $q$ render the original system solvable. Since $-\pi/2<\arctan<\pi/2$, from Eq. \ref{eq: continuity} we find
\begin{equation}
    -\pi<\am \left[ \frac{k}{2\xi_c}\left(d- q X_k\right)\right]<\pi.
\end{equation}
Finally, we note that $\am \left[\pm \frac{k X_k}{\xi_c}\vert -k^{-2}\right] = \pm \pi$, and  that the Jacobi Amplitude function is injective, then $q$ must satisfy
\begin{equation}
    \frac{d}{X_k}-2<q<\frac{d}{X_k}+2.
\label{eq: q cotes AP}
\end{equation}

\bibliography{bibliography,mybibliography}

\end{document}